\documentclass[iop]{emulateapj}

\newcommand{\myemail}{turrutia@aip.de}

\slugcomment{To appear in ApJ, hopefully}

\shorttitle{Spitzer observations of young red quasars}
\shortauthors{Urrutia et al.}

\begin{document}

\title{Spitzer observations of young red quasars}

\author{Tanya Urrutia\altaffilmark{1}, 
Mark Lacy\altaffilmark{2},
Henrik Spoon\altaffilmark{3},
Eilat Glikman\altaffilmark{4},
Andreea Petric\altaffilmark{5},
Bernhard Schulz\altaffilmark{6}
}

\altaffiltext{1}{Leibniz Institut f\"ur Astrophysik Potsdam, An der 
Sternwarte 16, 14482, Potsdam, Germany; \myemail}

\altaffiltext{2}{NRAO, 520 Edgemont Road, Charlottesville, VA 22903; 
mlacy@nrao.edu}

\altaffiltext{3}{Cornell University, Department of Astronomy, 219 Space 
Sciences Building, Ithaca, NY 14853; spoon@isc.astro.cornell.edu}

\altaffiltext{4}{Department of Physics, Yale University, P.O. Box 208120, New 
Haven, CT 06520, eilat.glikman@yale.edu}

\altaffiltext{5}{Astronomy Department, California Institute of Technology, 
Pasadena, CA 91125; ap@astro.caltech.edu}

\altaffiltext{6}{Infrared Processing and Analysis Center, California 
Institute of Technology, MC 100-22, Pasadena, CA 91125; 
bschulz@ipac.caltech.edu}

\begin{abstract}
We present mid-infrared spectra and photometry of thirteen redshift $0.4<z<1$ 
dust-reddened quasars obtained with {\it Spitzer} IRS and MIPS. We 
compare properties derived from their infrared spectral energy distributions 
(intrinsic AGN luminosity and far-infrared luminosity from star formation) to 
the host luminosities and morphologies from HST imaging, and black hole 
masses estimated from optical and/or near-infrared spectroscopy. Our results 
are broadly consistent with models in which most dust reddened quasars are an 
intermediate phase between a merger-driven starburst triggering a completely 
obscured AGN, and a normal, unreddened quasar. We find that many of our 
objects have high accretion rates, close to the Eddington limit. These 
objects tend to fall below the black hole mass -- bulge luminosity relation 
as defined by local galaxies, whereas most of our low accretion rate objects 
are slightly above the local relation, as typical for normal quasars at these 
redshifts. Our observations are therefore most readily interpreted in a 
scenario in which galaxy stellar mass growth occurs first by about a factor 
of three in each merger/starburst event, followed sometime later by black 
hole growth by a similar amount. We do not, however, see any direct evidence 
for quasar feedback affecting star formation in our objects, for example in 
the form of a relationship between accretion rate and star formation. Five of 
our objects, however, do show evidence for outflows in the 
[\ion{O}{3}]5007\AA~ emission line profile, suggesting that the quasar 
activity is driving thermal winds in at least some members of our sample.
\end{abstract}


\section{Introduction}

The study of quasars across cosmic time has proven to be important as their 
evolution and growth is intimately linked to that of their host galaxy. 
Mergers have long been invoked as the primary ignition mechanism of quasar 
activity, at least for the high luminosity objects 
\citep{sanders96,redqso-hst,nicola}. The loss of angular momentum during a 
gas rich merger allows the gas to be funneled to the center igniting a 
starburst and fueling the central black hole. In this model, eventually the 
quasar grows strong enough and develops a feedback mechanism that expels the 
obscuring material and shuts off star formation \citep{silkrees,hopkins08}. 
This scenario is further supported by the existence of tight correlations 
between the properties of galaxy bulges and their central black holes 
\citep{magorrian,ferrarese}.

Putting the picture together observationally is challenging, however. 
Populations of quasars selected via different techniques, although 
overlapping, differ substantially in their detailed properties and may bias 
towards a particular type of quasar or evolutionary phase e.g. AGN selected 
in the infrared may belong to different class or evolutionary phase than AGN 
selected in the radio or optical regime \citep{hickox09}. In an ideal survey 
we would be able to select quasars at specific points in their lifetimes. For 
example, to test a starburst-AGN connection we would prefer to select young, 
recently ignited objects to see if there is still star formation occurring in 
the host galaxies, and to identify likely feedback mechanisms.

One approach to finding candidates for young quasars is to search for dust 
obscured quasars that may correspond to objects yet to fully clear out the 
dust and gas surrounding them. Conventionally, obscured quasars are divided 
into two classes: (1) the so-called type-2 quasars - objects showing only 
narrow emission lines with large inferred extinctions ($A_V > 5-100$) towards 
the nucleus, most likely caused by a torus surrounding the accretion disk as 
required in the AGN  ``unification by orientation'' model 
\citep[e.g.,][]{antonucci,urry} and (2) moderately reddened quasars 
($A_V \sim 1-5$), which still show broad emission lines in the rest-frame 
optical, and whose continuum is still dominated by the quasar rather than the 
host galaxy. These latter objects are obscured by a cold absorber along the 
line of sight to the quasar, most likely in the host galaxy. They may thus 
represent the young objects in the final stages of emerging from their dusty 
cocoons. Throughout this paper the terms red, dust-obscured and moderately 
reddened quasars all refer to the latter category of objects.

Dust obscured quasars of all types are missing from, or are severely 
underrepresented in optical surveys, but are present in the radio 
\citep[e.g.,][]{postman}, in the X-ray \citep[e.g.,][]{hickox07}, when 
selected on the basis of narrow optical AGN emission lines 
\citep{zakamska03}, or when selected in the near-infrared \citep{cutri}. 
Mid-infrared surveys 
\citep[e.g.,][]{lacy04,lacy07a,stern05,alonso06,donley07} have also been 
extremely successful at finding and identifying obscured quasars over a wide 
range of reddenings, redshifts, and luminosities. Surveys combining deep 
radio and mid-infrared data show that the optically- and/or X-ray-selected 
quasar population constitutes less than half of the total population of 
quasars \citep{donley05,martinez05}. Joint selection using radio and 
near-infrared, as used for the sample in this paper, has proven to be one of 
the most reliable ways to select the moderately obscured quasar population. 
The requirement of a bright radio source and very red optical through 
near-infrared colors results in a set of candidates with relatively little 
contamination from normal galaxies, low luminosity AGN and stars 
\citep{gregg02,lacy02,eilat04,f2m07,glikman12,f2ms}.

The success of the {\it Spitzer Space Telescope} \citep{spitzer} 
has allowed for the detailed study of quasars in the mid-infrared. Programs 
with the Infrared Spectrograph \citep[IRS;][]{irs}, in particular, have 
refined our knowledge of QSO spectral energy distributions (SEDs) in the 
mid-infrared. For example, the detection of silicate in emission in IRS 
spectra of quasars is a strong support for the unification model 
\citep{hao05,siebenmorgen05}. Average properties of classes of objects using 
IRS spectra \citep[e.g,][]{spoon07} have allowed us to constrain the physical 
properties of dust in AGN \citep[e.g.,][]{nikutta09}. On average the spectra 
of luminous quasars are flat, show little or no PAH emission and the silicate 
features are in emission, associated with dust re-emission. In contrast, the 
spectra of Ultraluminous Infrared Galaxies (ULIRGs) show a steep rise towards 
the long wavelengths, moderate PAH emission and silicate absorption troughs 
associated with embedded star formation \citep{hao07}. However, some type 2 
and some reddened quasars at moderate to high redshifts show deep silicate 
absorption features \citep{lacy07b,martinez08,zakamska08} in contrast to 
samples of X-ray selected Type-2 quasars, in which they are generally weak 
\citep{sturm06}. 

The Multiband Imaging Photometer for SIRTF \citep[MIPS;][]{mips} on board 
{\it Spitzer} has also allowed us to sample the colder part of the galaxy 
SED. While AGN dominate the bright 24$\mu$m population \citep{donley08}, the 
{\it Spitzer} 70$\mu$m field population is dominated by ULIRGs which 
have a high merger fraction \citep{kar10a,kar10b}. However, AGN and quasars 
still make up a significant fraction of 70$\mu$m sources, and high star 
formation rates in quasar hosts are common \citep[e.g.][]{quest1,quest2}. 
Consistent with those results \cite{shi09} find that Type-1 quasars show star 
formation rates higher than the field galaxies, with luminosities typical of 
Luminous Infrared Galaxies (LIRGs, $L_{IR} = 10^{11-12} L_{\odot}$). In 
particular, dust reddened  quasars have been found to have higher than usual 
60/12$\mu$m luminosity ratios, an indicator of higher star formation rates 
\citep{georgakakis}.

In this paper we will describe {\em Spitzer} IRS and MIPS observations 
of 13 redshift $0.4<z<1$ reddened quasars selected using joint radio and 
near-infrared selection for which we have also obtained Hubble Space 
Telescope (HST) observations \citep{redqso-hst}. We use these observations to 
estimate the intrinsic luminosities of the quasars, and to estimate or 
constrain the star formation rates in the host galaxies. We then evaluate the 
evidence that this population corresponds to an intermediate stage in quasar 
evolution by estimating black hole masses and accretion rates, comparing our 
HST host galaxy luminosities and morphologies to the other properties, and by 
searching for evidence of feedback processes at work in the hosts. Throughout 
this paper we adopt a flat Universe, $H_0$ = 70 km s$^{-1}$ Mpc$^{-1}$, 
$\Omega_{\Lambda} = 0.7$ cosmology.

\begin{deluxetable*}{lccccc}
\tabletypesize\scriptsize
\tablecaption{{\it Spitzer} IRS and MIPS observations 
\label{observe}}
\tablewidth{0pt}
\tablehead{
\colhead{Source} & \colhead{IRS AORID} & \colhead{MIPS AORID} & \colhead{$S_{24}$ (mJy)} & \colhead{$S_{70}$ (mJy)} & \colhead{$S_{160}$ (mJy)}}
\startdata
F2M0729$+$3336 &22386176 &22386432 &12.9$\pm$3.0 &$<$15       &$<$30  \\
F2M0825$+$4716 &22389760 &22386688 &36.1$\pm$4.7 &85$\pm$19   & 46$\pm$11  \\
F2M0830$+$3759 &22390016 &22386944 &26.1$\pm$3.9 &32$\pm$12   & $<$30  \\
F2M0834$+$3506 &22390272 &22387200 &17.6$\pm$3.4 &26$\pm$13   & $<$30  \\
F2M0841$+$3604 &22390528 &22390528 &11.7$\pm$3.0 &53$\pm$14   & $<$30  \\
F2M0915$-$2418 &17540352$^{\dag}$ &22387712 & 87.3$\pm$7.0 &207$\pm$24 &73$\pm$12 \\
F2M1012$+$2825 &22390784 &22387968 & 4.9$\pm$0.8 &$<$15       & $<$30 \\
F2M1113$+$1244 &22391040 &22388224 &59.4$\pm$5.8 &138$\pm$25  & 67$\pm$12 \\
F2M1118$-$0033 &22391296 &22388480 &24.4$\pm$4.0 &80$\pm$24   & 44$\pm$10 \\
F2M1151$+$5359 &22391552 &22388736 &6.9$\pm$0.7  &20$\pm$7    & $<$30 \\
F2M1507$+$3129 &22391808 &22388992 &9.2$\pm$2.5  &22$\pm$11   & $<$30 \\
F2M1532$+$2415 &22392064 &22389248 &28.4$\pm$4.0 &31$\pm$13   & $<$30 \\
F2M1656$+$3821 &22392320 &22389504 &9.7$\pm$2.4  &15$\pm$8    & $<$30 \\
\enddata
\tablecomments{All observations made in cycle 4 (Jul 2007 - May 2008) except
for F2M0915-2418, whose IRS spectrum was taken in May 2007 as part of program
30121.}
\end{deluxetable*}

\section{Quasar sample and observations}

In the past few years we have been selecting a sample of luminous 
lightly-obscured quasars using a combination of the FIRST radio survey 
\citep{first}, the 2MASS near infrared survey \citep{2mass} and the Cambridge 
APM scans of the POSS plates. Using a $J-K > 1.7$ and $R-K > 4$ color wedge 
for the matched point sources, we have been very effective in finding these 
so-called red quasars with spectroscopic follow up at Keck (ESI) and IRTF 
(Spex) \citep{eilat04,f2m07,f2ms}. Even though the survey is radio selected, 
most of the objects rather fall in the radio intermediate regime than 
radio-loud, so it is unlikely that the redness of the objects comes from a 
synchrotron component, but is in fact associated with dust-reddening. 
Furthermore, the excess of steep spectrum sources at faint radio fluxes and 
the point-source nature of the radio sources leads us to expect that the 
nature of the radio emission is due to a quasar and not from star formation 
\citep{f2m07}. As of early 2012, we have well over 120 spectroscopically 
confirmed red quasars found with this method \citep{glikman12}.

\begin{figure}
\begin{center}
\plotone{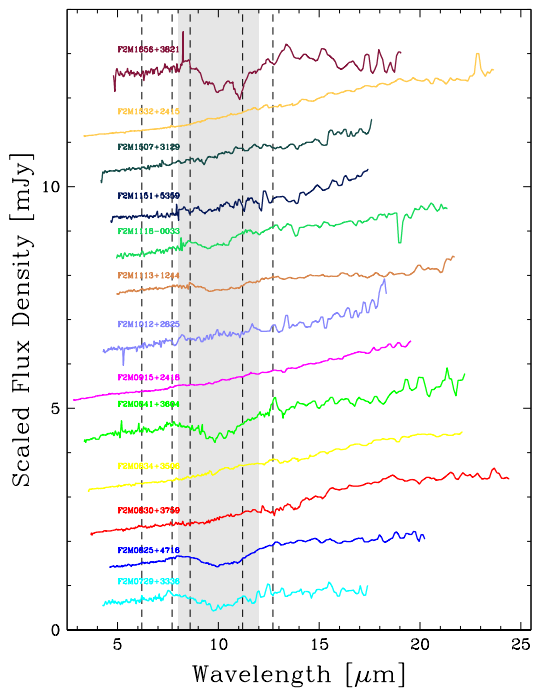}
\caption{Stacked IRS rest-frame spectra of red quasars. The spectra are 
normalized to 1 at 15 $\mu$m and offset. The dashed lines denote the central 
wavelengths of the family of PAH emission features at 6.2, 7.7, 8.6, 11.2, 
and 12.7 \micron. The shaded region represents the typical region (8.0-12.0 
\micron) where the 9.7 \micron~ silicate absorption feature is seen.}
\label{irsspec}
\end{center}
\end{figure}

These red quasars are highly luminous and only moderately reddened 
($A_V\sim 1-5$ mags), so they outshine their host galaxies in the rest-frame 
optical, and are thus not red due to galaxy starlight. Their spectral energy 
distributions (SEDs) are generally well fit by reddening a standard quasar 
template spectrum using an SMC dust-reddening law. It is only in the 
rest-frame UV, where the quasar light is heavily extinguished, that the host 
galaxy starts to play a role in terms of overall flux density. See 
\cite{f2m07,f2ms} for in depth discussions of how reddening is estimated
for these objects.

A representative sample of 13 dust-reddened quasars with $0.4<z<1$ were 
followed up with the ACS Wide Field Camera on HST with the F475W and F814W 
filters. These are very luminous quasars at $z \sim 0.7$ having intrinsic 
absolute magnitudes in the range $-23.5 \ge M_B \ge -26.2$. The host galaxies 
show a high amount of interaction: 85\% of the images show evidence of 
morphological disturbance \citep{redqso-hst}. We also observed that the more 
reddened the objects were, the more disturbed their morphologies (as measured 
by the Gini coefficient combined with the Concentration Index 
\citep{gini,cas}). These results support the merger-induced origin for high 
luminosity quasars.

We followed up this sample of 13 red quasars with the MIPS and IRS on board 
the {\it Spitzer Space Telescope} (PID 40143). The IRS 
observations were made with the Short-Low (SL; $\Delta\lambda$ =  5.2-14.5 
\micron) and Long-Low (LL; $\Delta\lambda$ = 14.0-38.0 \micron) modules of 
the IRS, (with some of the highest redshift objects not using the SL2 5-7.5 
\micron~module). We also obtained MIPS 24$\mu$m, 70$\mu$m and 160$\mu$m 
photometry (with the lowest redshift objects not being observed at 160 
$\mu$m). Table \ref{observe} provides the details for our Spitzer 
observations, including the IRS and MIPS AORIDs as well as the MIPS 24, 70, 
and 160 $\mu$m fluxes.

Our infrared spectra and photometry were used for a number of purposes. We 
can infer star formation rates from the PAH emission in the IRS spectra and 
the far-infrared excesses from the MIPS photometry. The silicate feature at 
9.7 $\mu$m provides us with an estimate of the cold dust obscuration, and the 
mid-infrared continuum emission provides an estimate of the intrinsic 
(unreddened) quasar luminosity. Near-infrared and optical spectra of these 
objects are shown in \cite{eilat04,f2m07}, though we obtained an additional 
spectrum of FTM1507+3129, see Section 4.1.11.

\begin{figure*}
\begin{center}
\includegraphics[width=13cm,angle=-90]{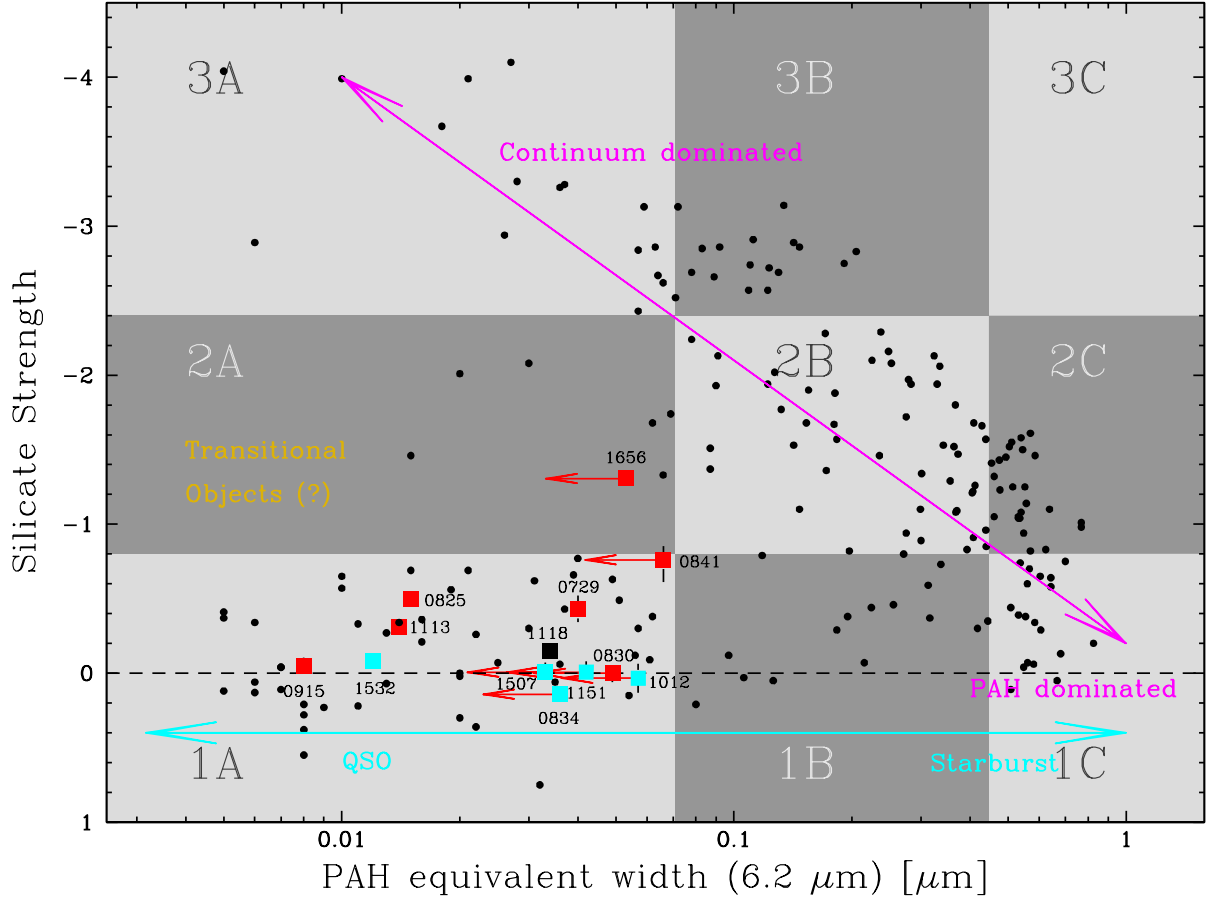}
\caption{The diagnostic diagram of \cite{spoon07} with their sources plotted 
as small filled circles. Our quasars are plotted as larger squares with the 
PAH values showing arrows being upper limits. Usually sources cluster along 
two tracks: the diagonal continuum track from 1C to 3A (magenta) spanning 
from Starburst to ULIRG-type systems and the activity track from 1A to 1C 
(cyan) spanning from QSOs to Starburst galaxies. Most of the quasars are 
found in quadrant 1A, but some have Silicate troughs deep enough to be in 
quadrant 2A, which \cite{spoon07} speculates to be transitional objects.}
\label{diagnostics}
\end{center}
\end{figure*}

\section{Data Reduction and Analysis}

The IRS spectra were reduced in the following manner. First we removed the 
bad or hot pixels using the IDL program \verb irsclean. The resulting spectra 
were background subtracted by differencing the first and second-order 
apertures. They were then extracted and flux calibrated using SPICE version 
2.0.1 provided by the {\it Spitzer} Science Center. Finally, the 
different orders were stitched together using a weighted average in the 
overlapping orders, with no significant scaling adjustments needed in the 
order-to-order jumps. A plot of the 13 different normalized spectra can be 
seen in Figure \ref{irsspec}.

We used the post-BCD products for the 24 \micron~MIPS images, as the quality 
was excellent and there was no need to re-mosaic them. For the 70 and 160 
\micron~data we re-mosaiced the data using \verb MOPEX  version 18.2.2 with 
the default parameters. We employed two methods for the photometric 
measurements: (a) The \verb APEX  package within \verb MOPEX  with the 
default parameters and PRFs for point-source photometry and (b) 
\verb SExtractor  aperture photometry with standard point-source aperture 
corrections provided by the Spitzer Science Center\footnote{Using the
old webpage http://ssc.spitzer.catech.edu/mips/apercorr, 
values can now be found in the MIPS instrument handbook at 
http://ssc.spitzer.caltech.edu/mips/mipsinstrumenthandbook/50/}. We used 10\% 
errors for the 70 and 160 \micron~sources, as \verb MOPEX  tended to 
underestimate the errors.

\subsection{IRS spectral analysis}

For all the IRS spectra in our sample we measured the equivalent width 
(EW) of the 6.2 \micron~PAH emission as well as the strength of the 9.7 
\micron~silicate absorption feature following the methodology presented in 
\cite{spoon07}. In short the PAH EW is obtained by integrating above a spline 
interpolated continuum, while the silicate feature strength is inferred from 
the ratio between the observed and continuum flux density:
\begin{equation}
S_{\rm Sil} = {\rm ln} \frac{f_{obs} (\mathrm{9.7 }\, \mu m)}{f_{cont} 
(\mathrm{9.7 }\, \mu m)}
\end{equation}
The values are given in Table \ref{fitting} and plotted in Figure 
\ref{diagnostics}. We discuss the results in the diagram in Section 4.

\begin{figure*}
\begin{center}
\hspace*{0cm}
\vspace*{0.5cm}
\includegraphics[height=2.87cm]{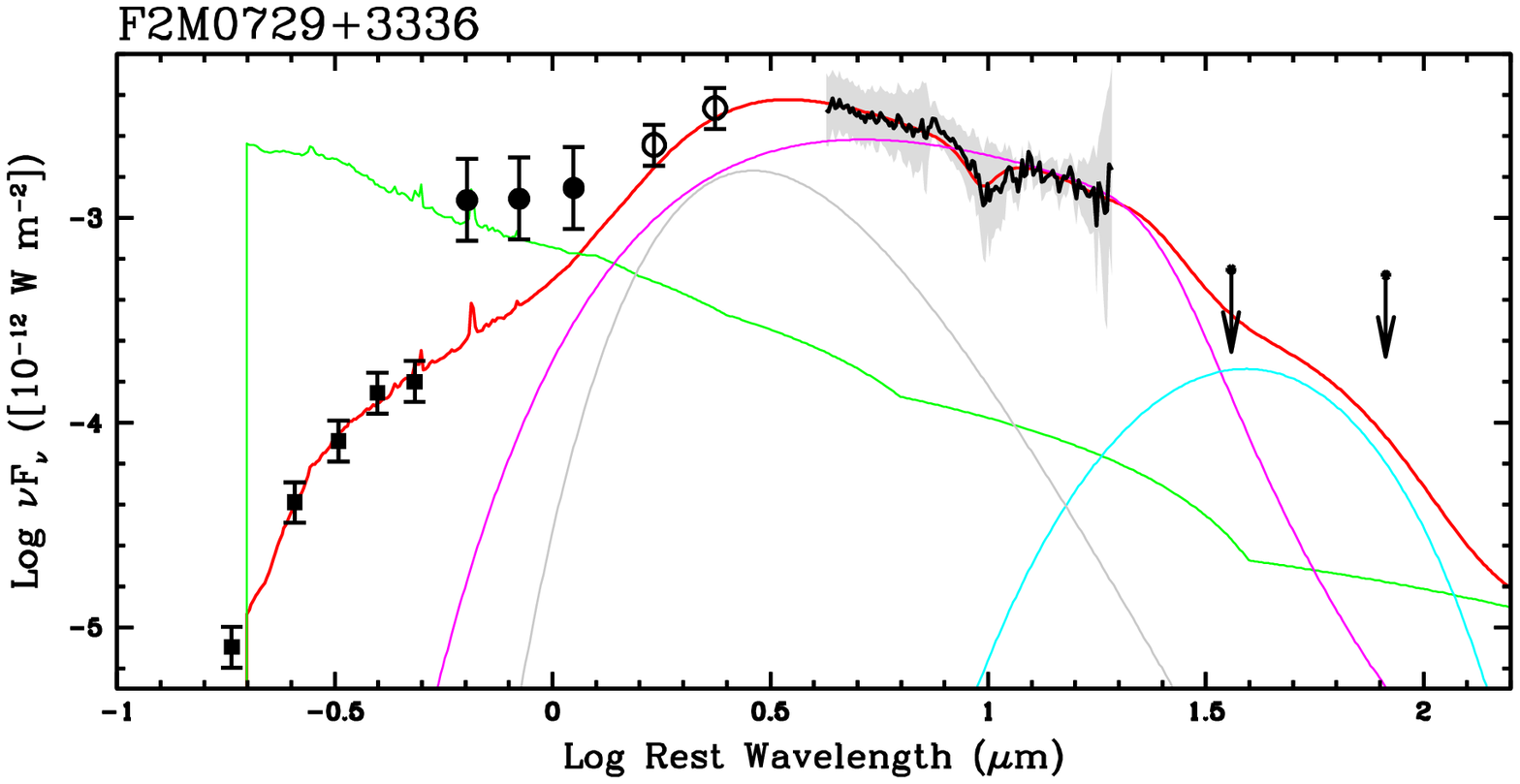}
\includegraphics[width=2.38cm]{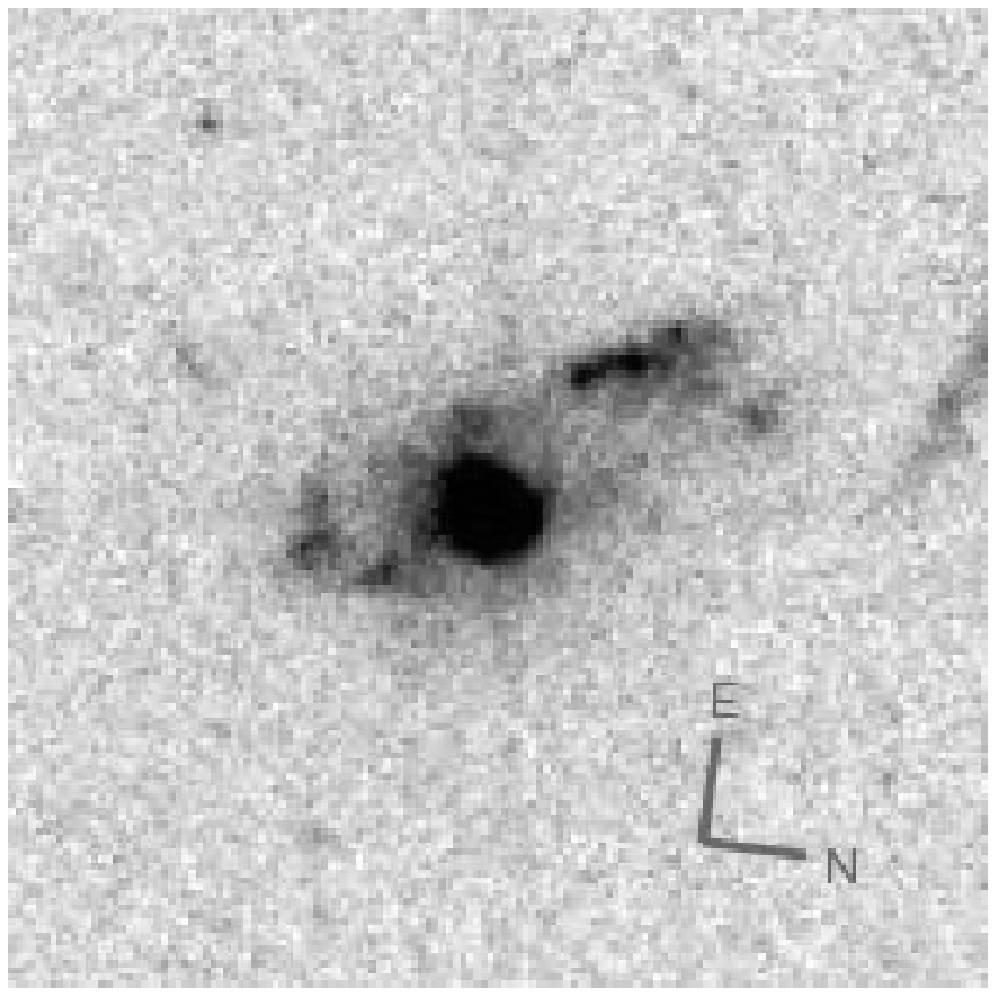}
\hspace*{0.1cm}
\includegraphics[height=2.88cm]{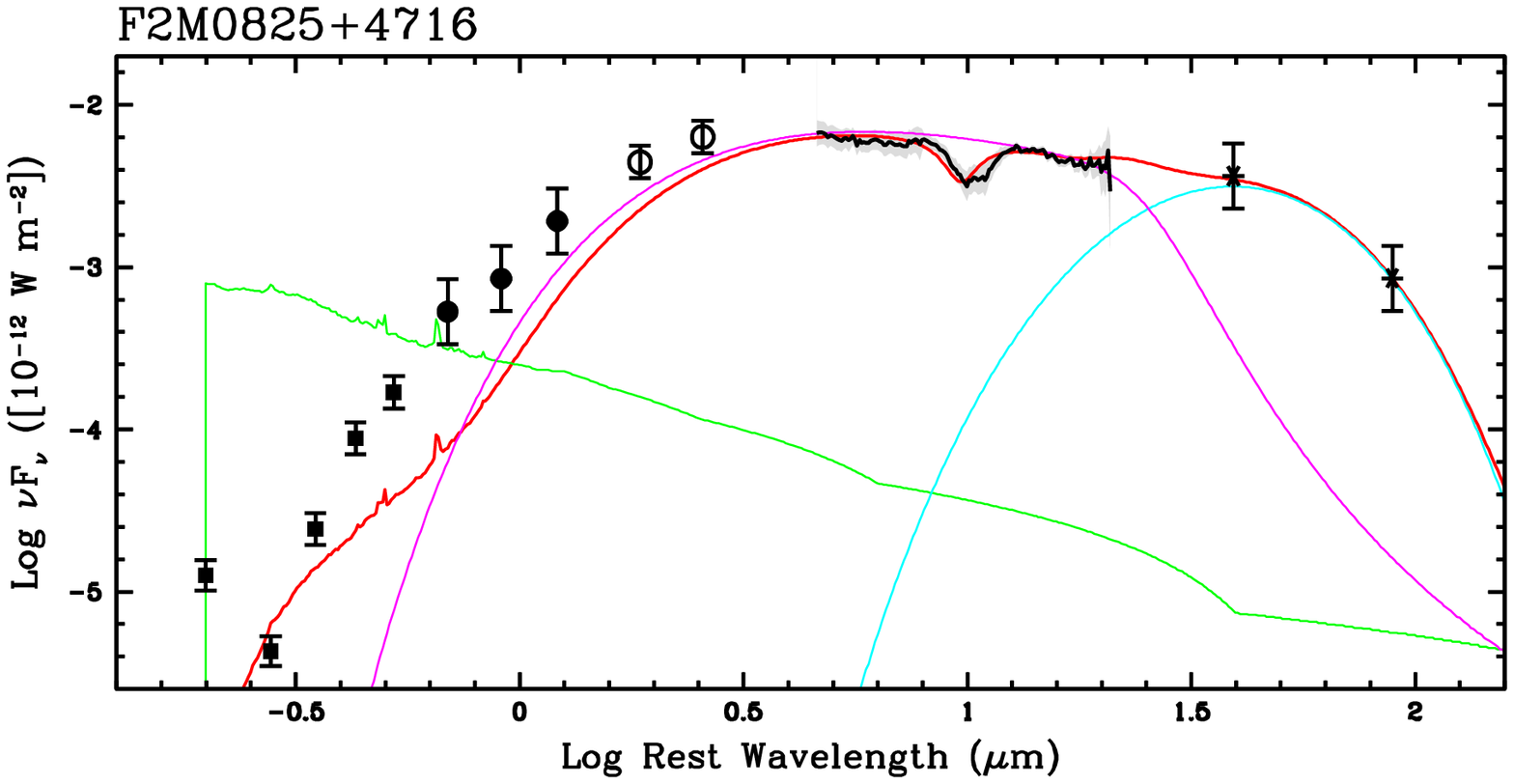}
\includegraphics[width=2.38cm]{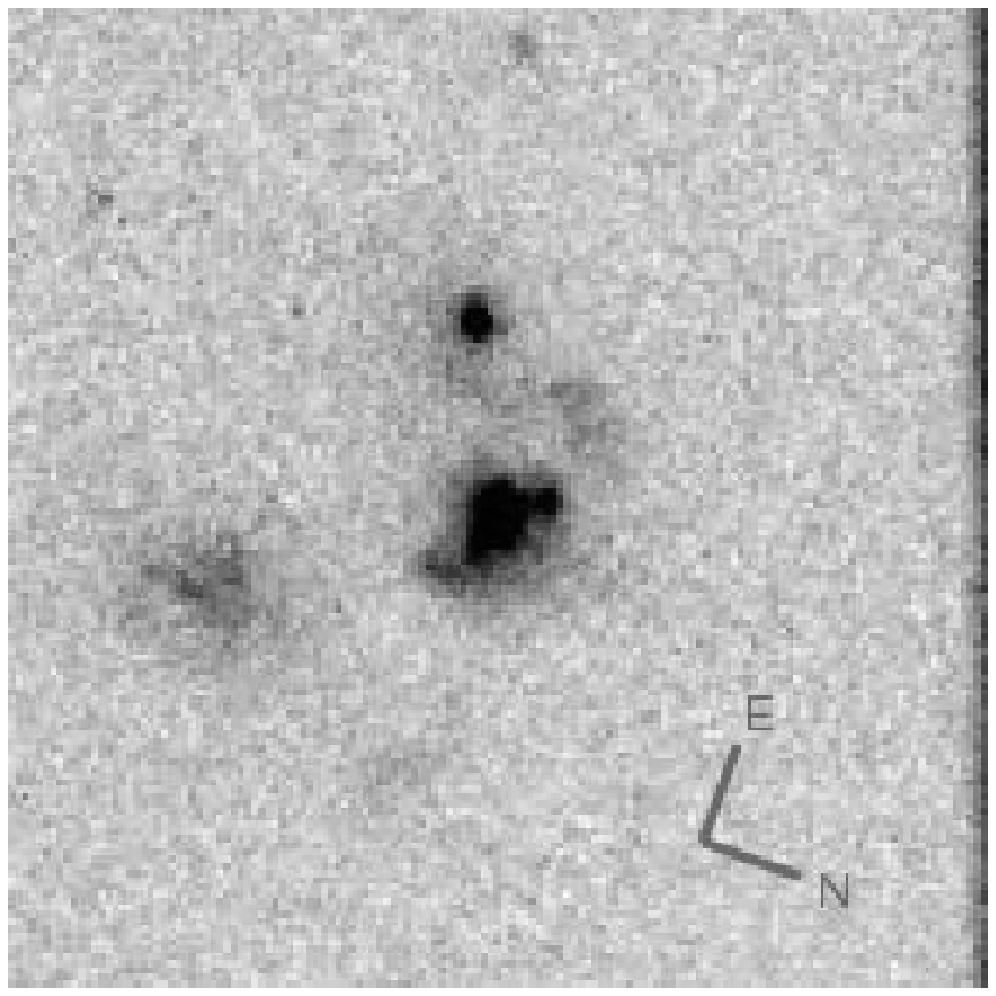}
\includegraphics[height=2.87cm]{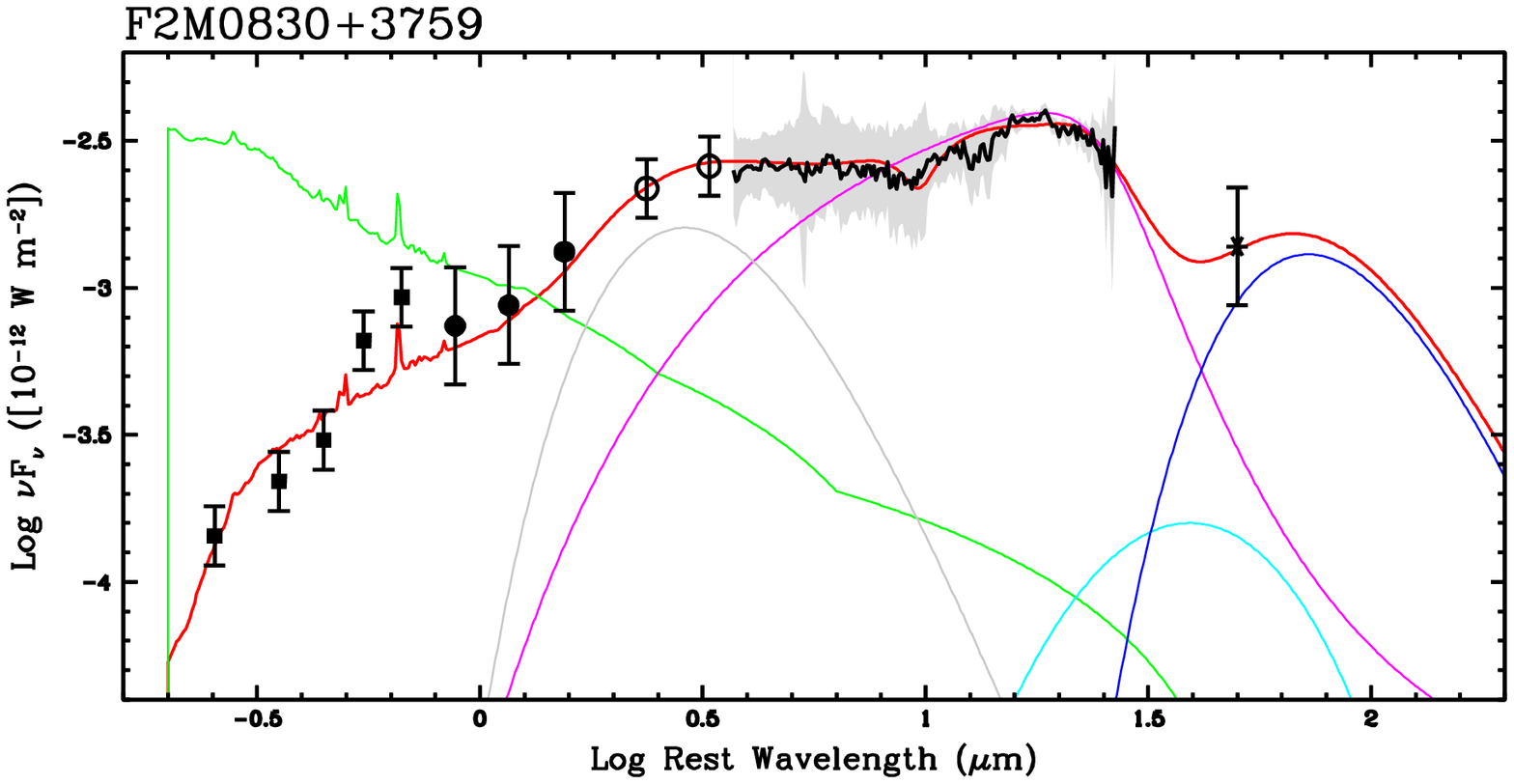}
\includegraphics[width=2.38cm]{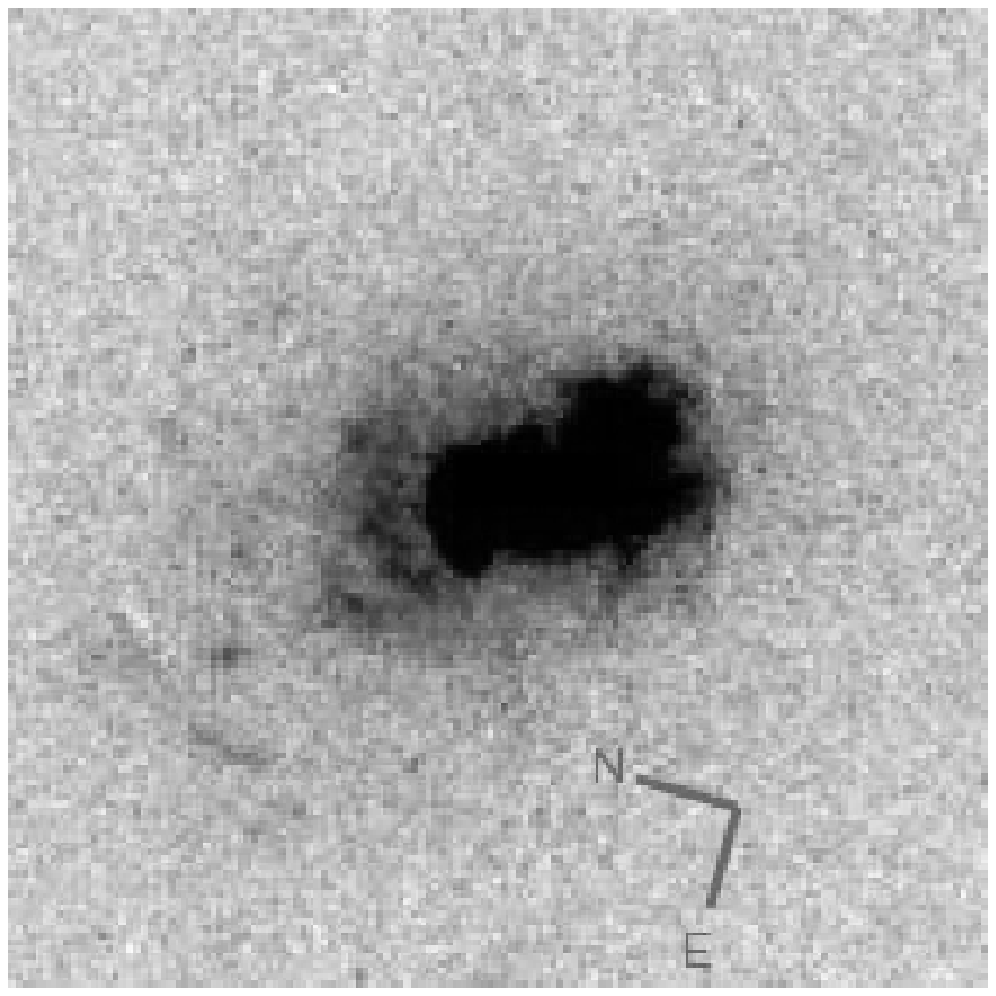}
\hspace*{0.1cm}
\vspace*{0.5cm}
\includegraphics[height=2.87cm]{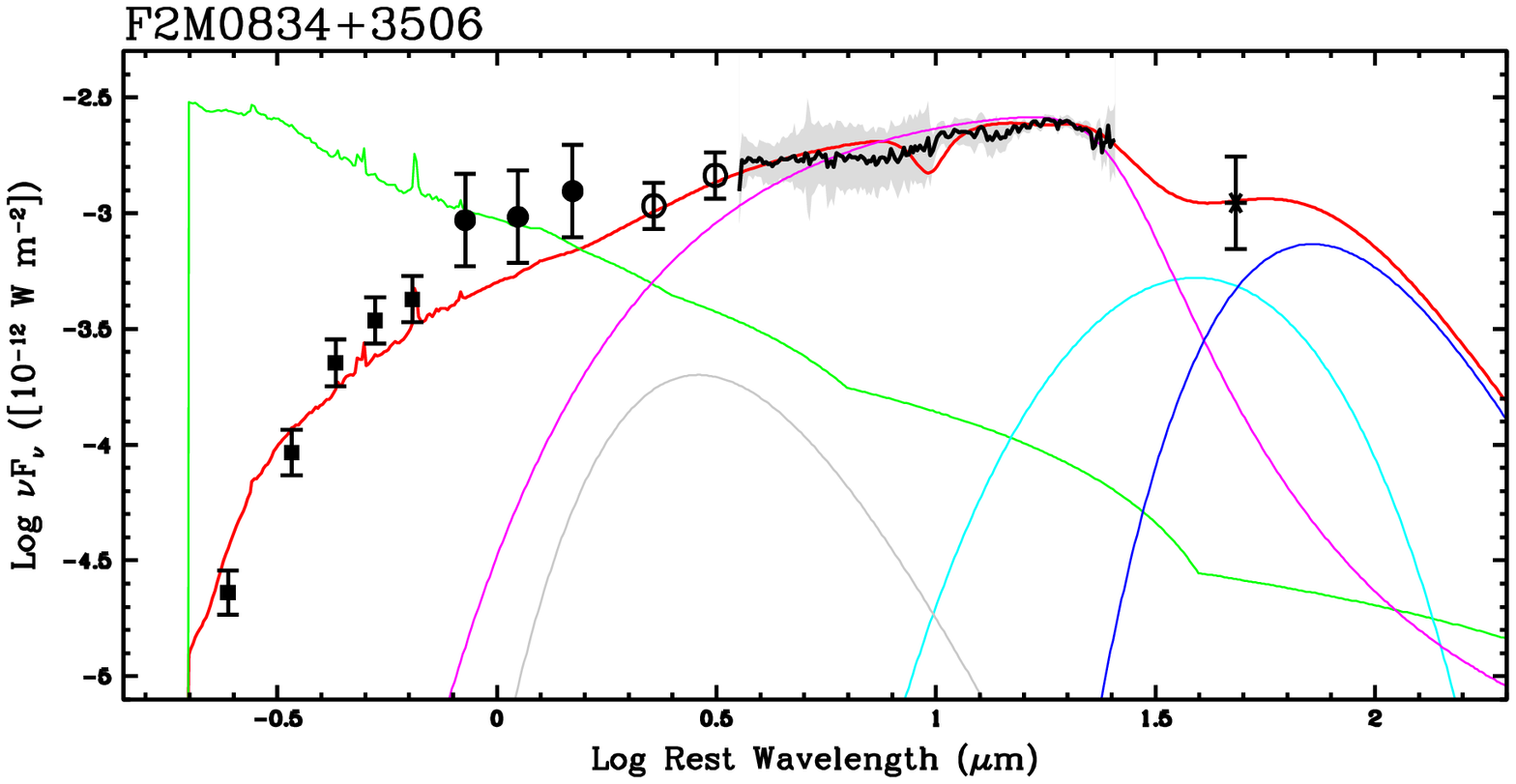}
\includegraphics[width=2.38cm]{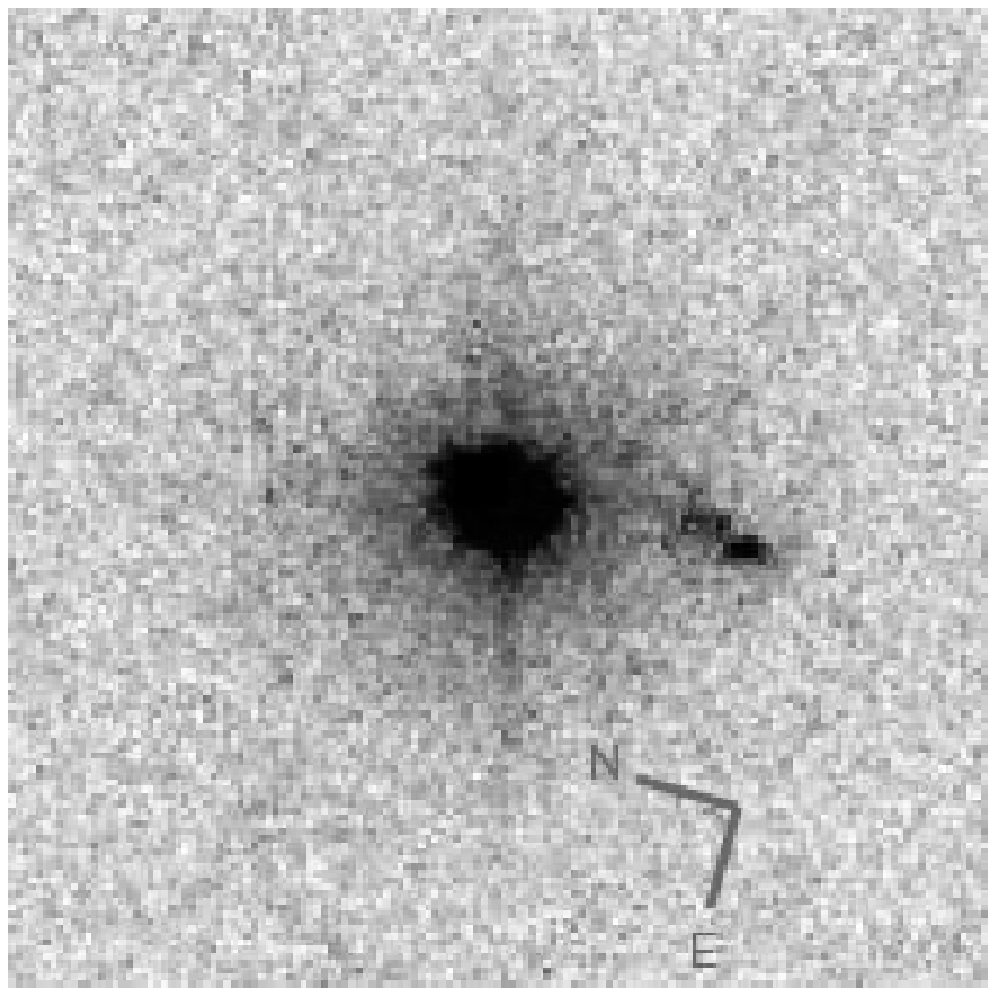}
\includegraphics[height=2.87cm]{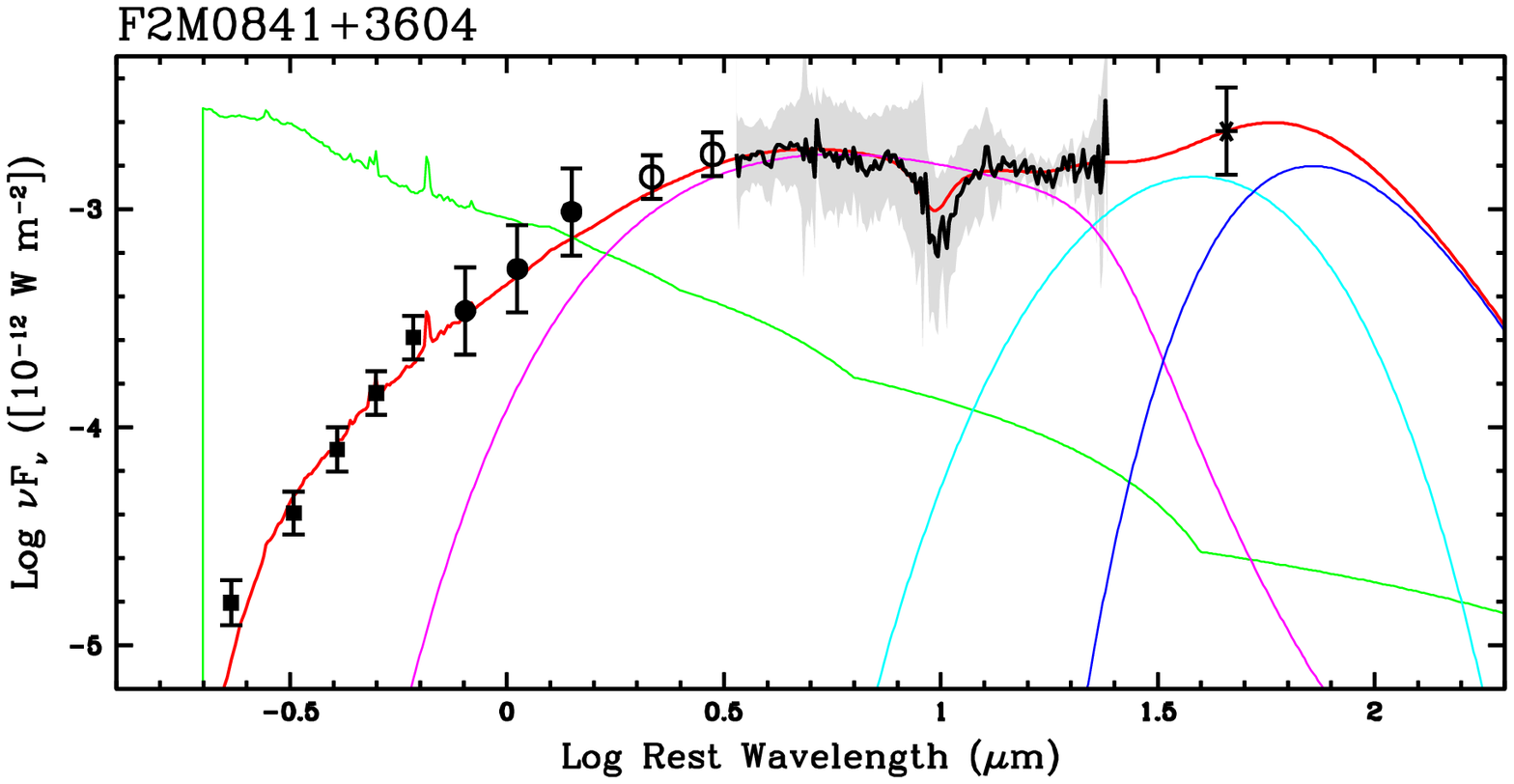}
\includegraphics[width=2.38cm]{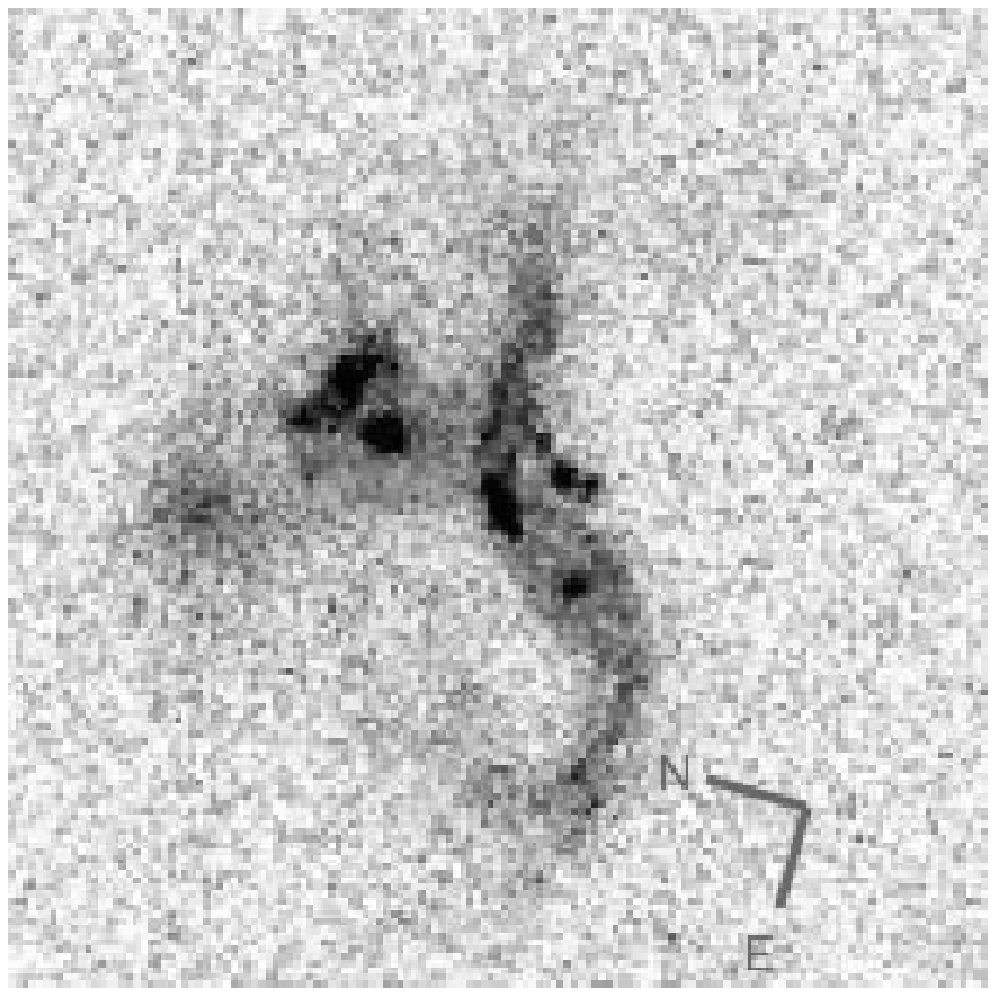}
\hspace*{0.1cm}
\vspace*{0.5cm}
\includegraphics[height=2.87cm]{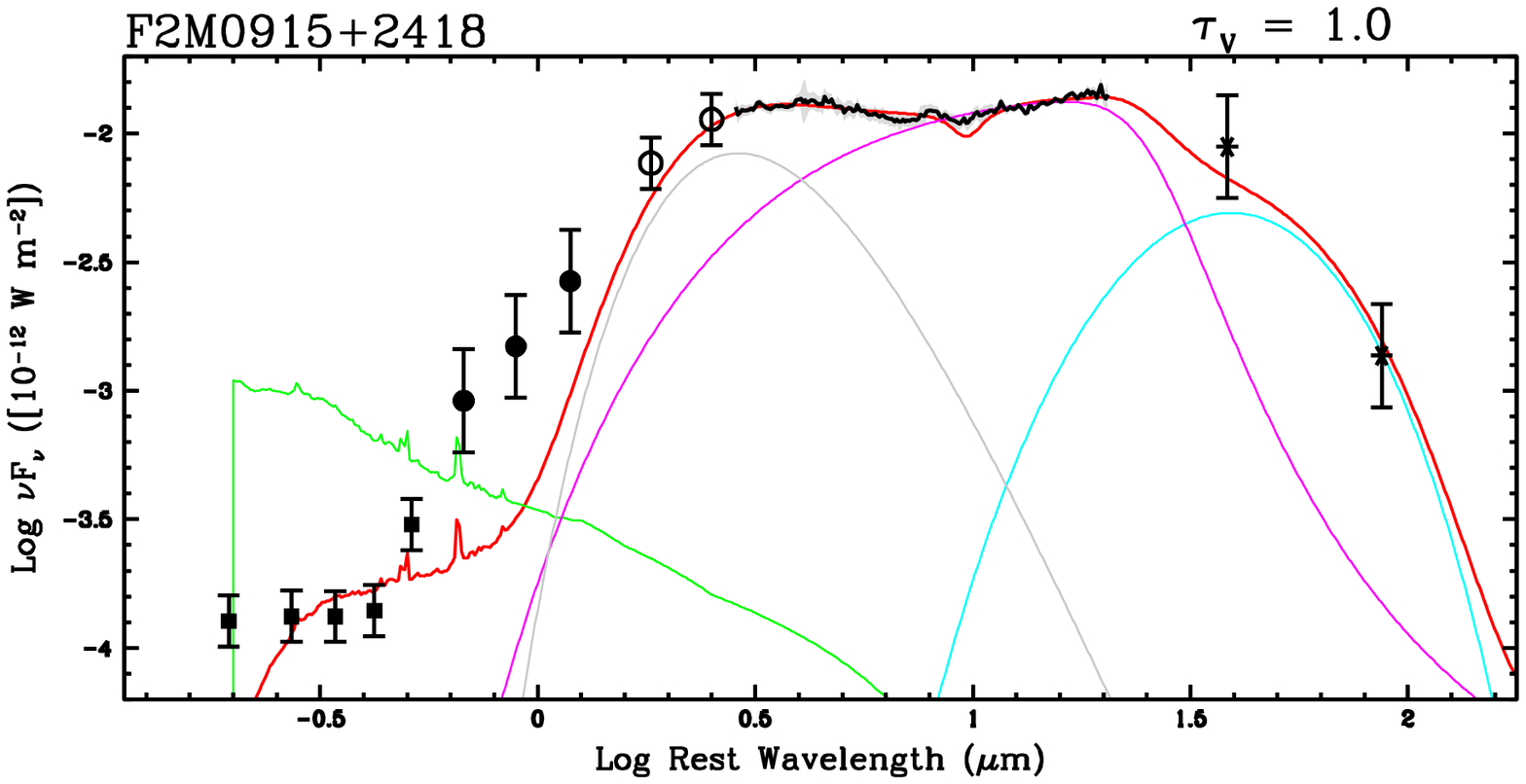}
\includegraphics[width=2.38cm]{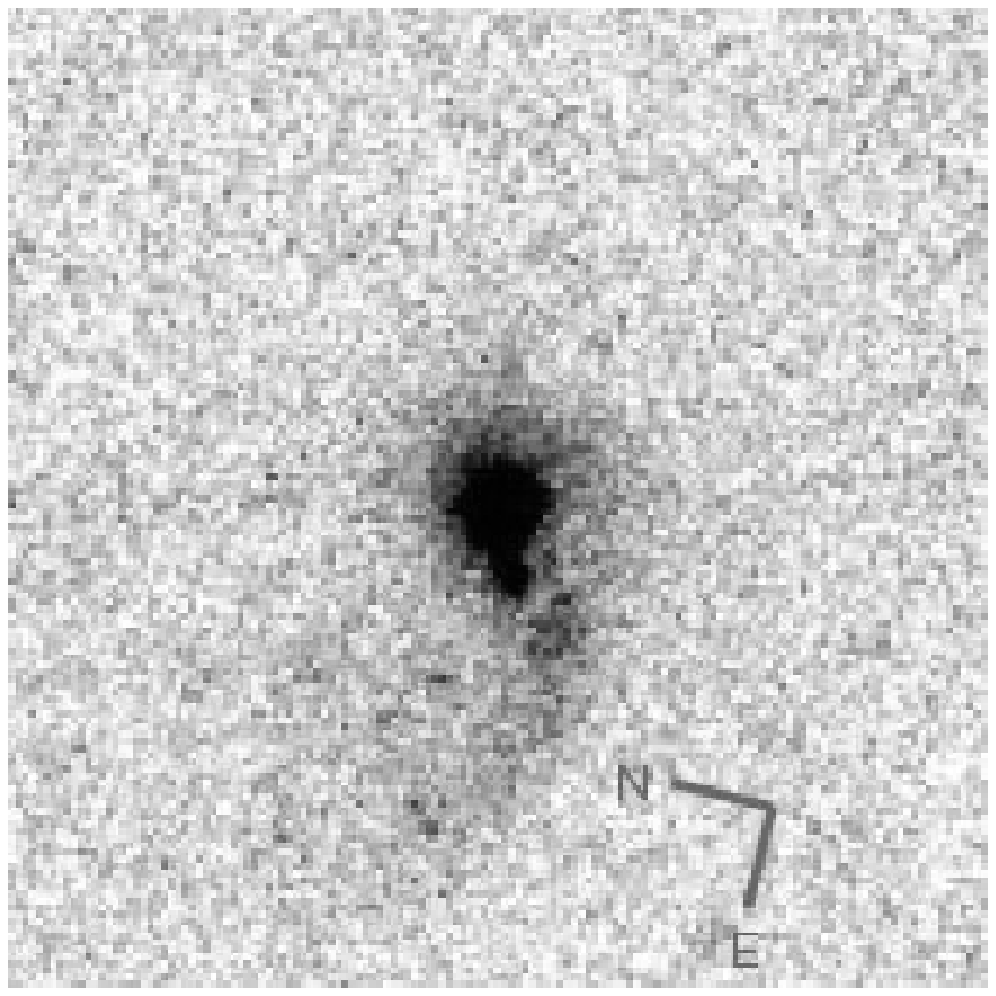}
\includegraphics[height=2.87cm]{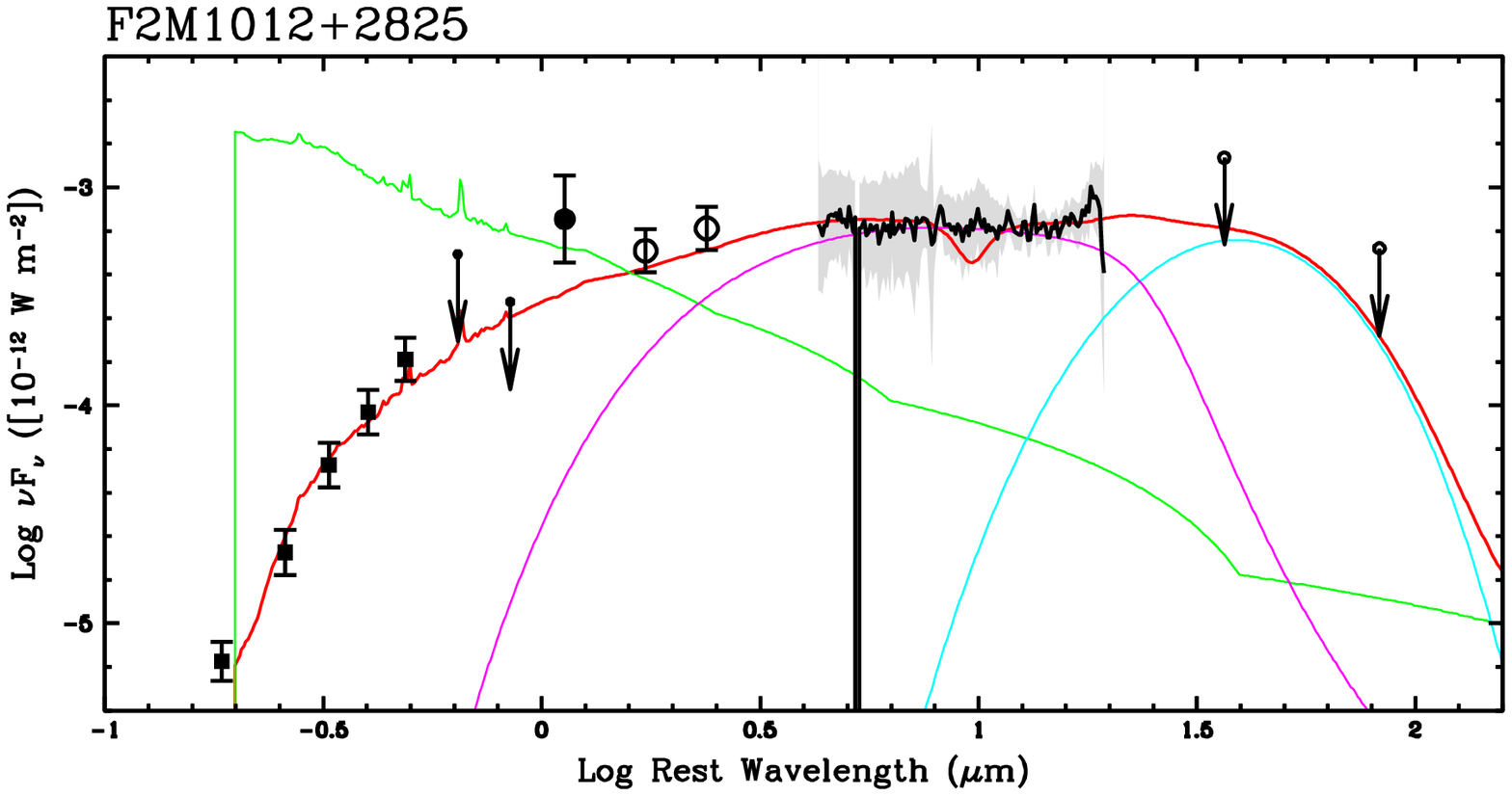}
\includegraphics[width=2.38cm]{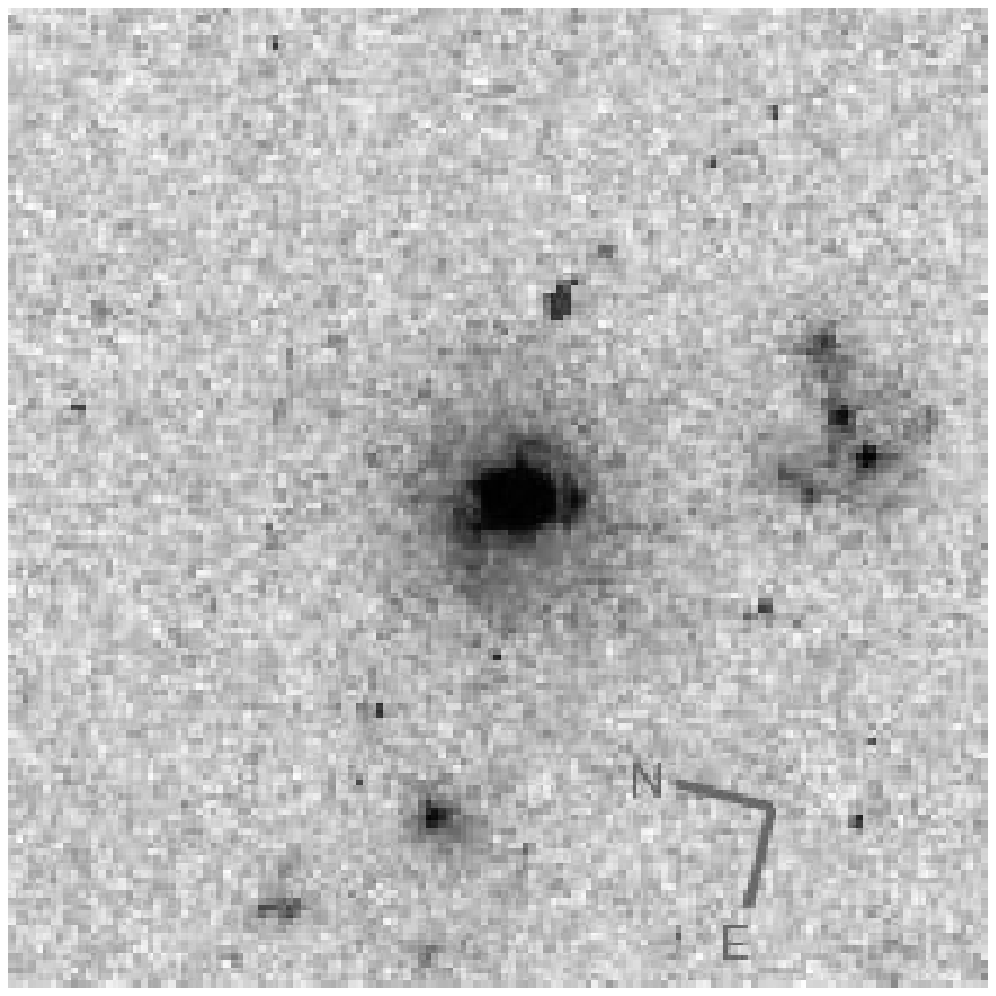}
\hspace*{0.1cm}
\vspace*{0.5cm}
\includegraphics[height=2.87cm]{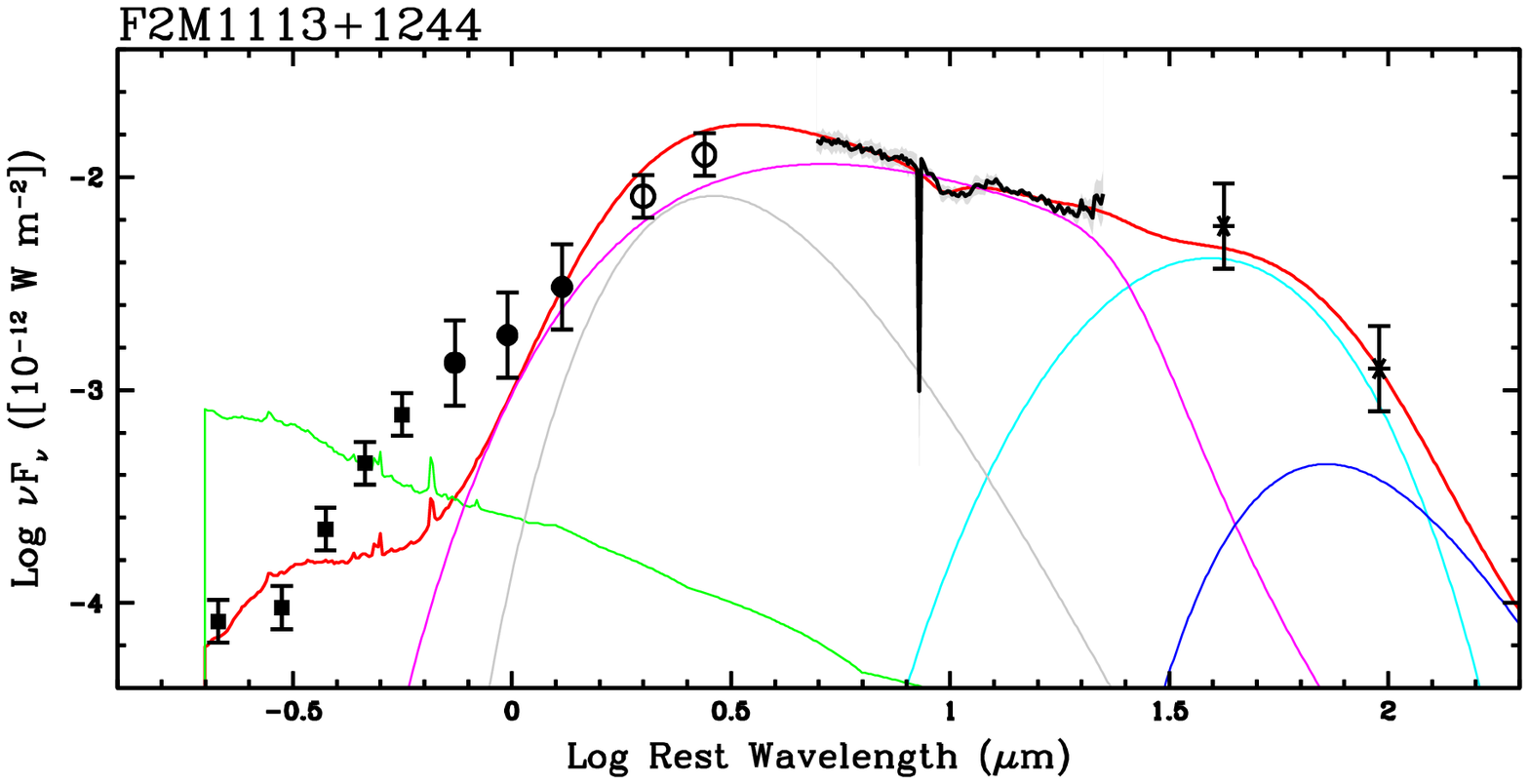}
\includegraphics[width=2.38cm]{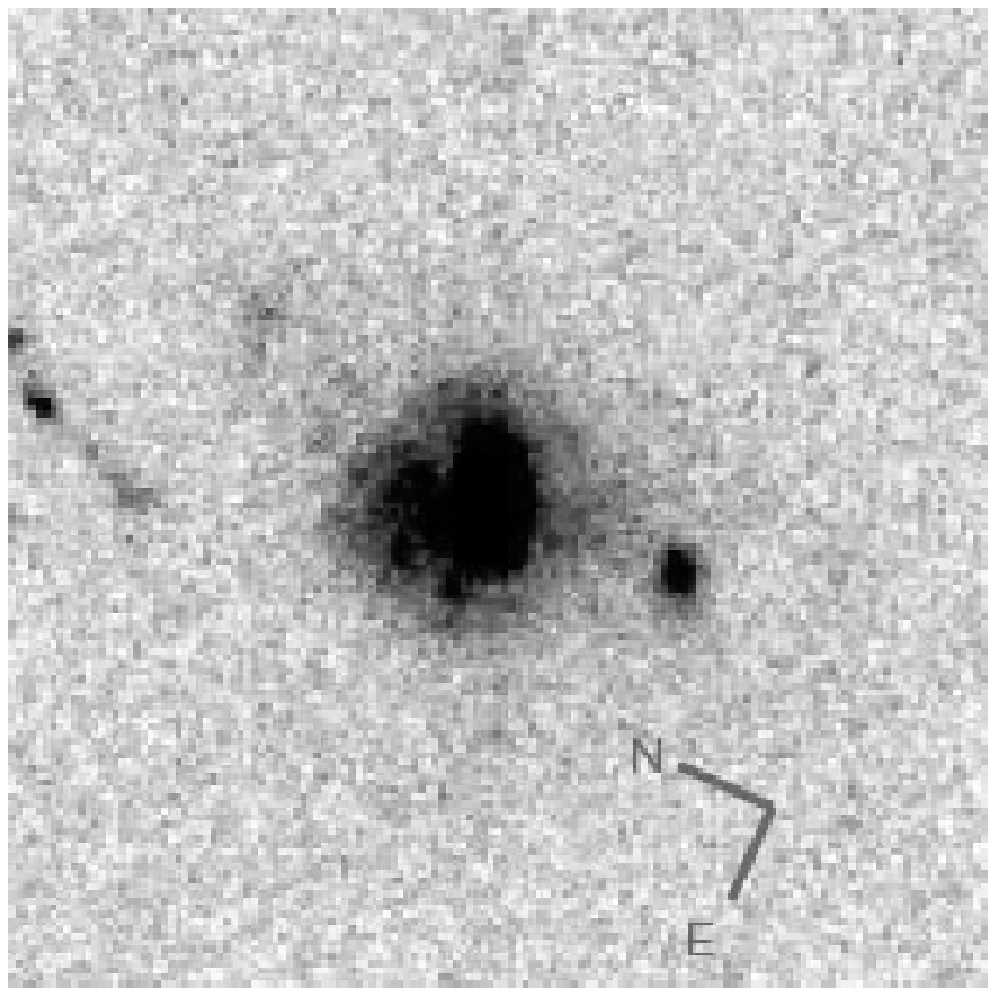}
\caption{Broadband rest-frame spectral energy distributions (SEDs) of the red 
quasars. The five photometry measurements from the SDSS are shown as squares, 
the ones from 2MASS as filled circles, the ones from WISE Band 1+2 as open 
circles and the ones from MIPS as stars (three of the low redshift quasars 
were not observed at MIPS 160\micron). Also shown is the IRS spectrum with a 
corresponding grey error area. The fitted SEDs of the quasars are shown in 
red. They are built up from a quasar component (in green) and AGN hot dust 
component (pink), both of which are reddened by an SMC extinction law. 
Furthermore a very hot quasar dust component (grey), a warm dust component 
(cyan) and a cold dust component (blue) are added. The corresponding HST 
image is shown to the right of each SED as an $7\times7$ arcsec greyscale 
image. }
\label{comparison}
\end{center}
\end{figure*}

\subsection{Multiwavelength Modeling}\label{model}

To obtain a complete view of the physical phenomena that are affecting the 
quasar, we modeled the SEDs of our sample from the optical out to the 
mid-infrared. We used the available $u,g,r,i,z$-band SDSS 
photometry\footnote{Except for F2M0729$+$3336, which has no SDSS coverage. We 
obtained synthetic photometry corresponding to the SDSS filters from the 
optical spectrum} in the optical. In addition, we used 2MASS \citep{2mass}, 
WISE Band 1+2 \citep{wise}, our MIPS photometry and IRS mid-IR spectra.

\setcounter{figure}{2}
\begin{figure*}
\begin{center}
\hspace*{0cm}
\vspace*{0.5cm}
\includegraphics[height=2.87cm]{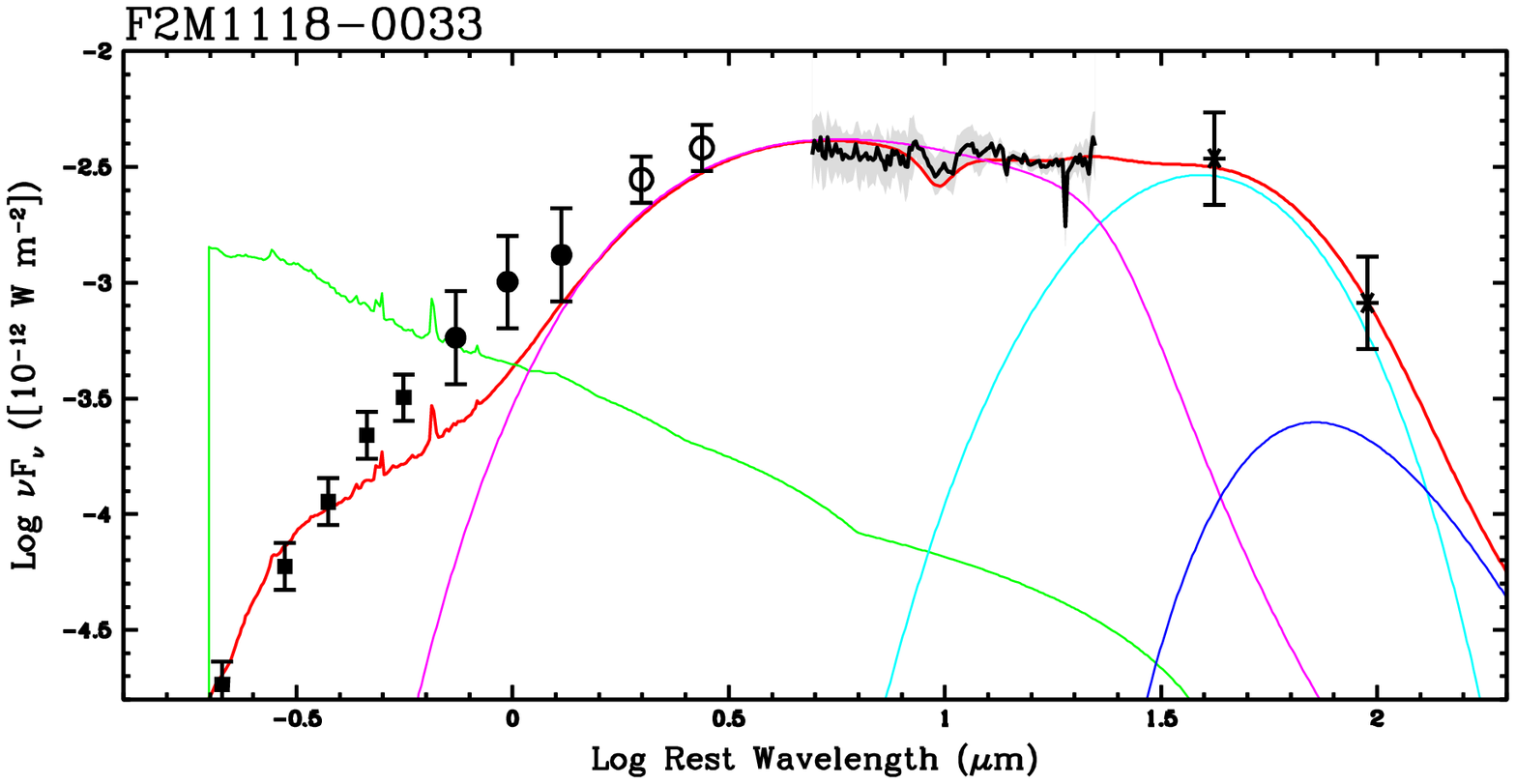}
\includegraphics[width=2.38cm]{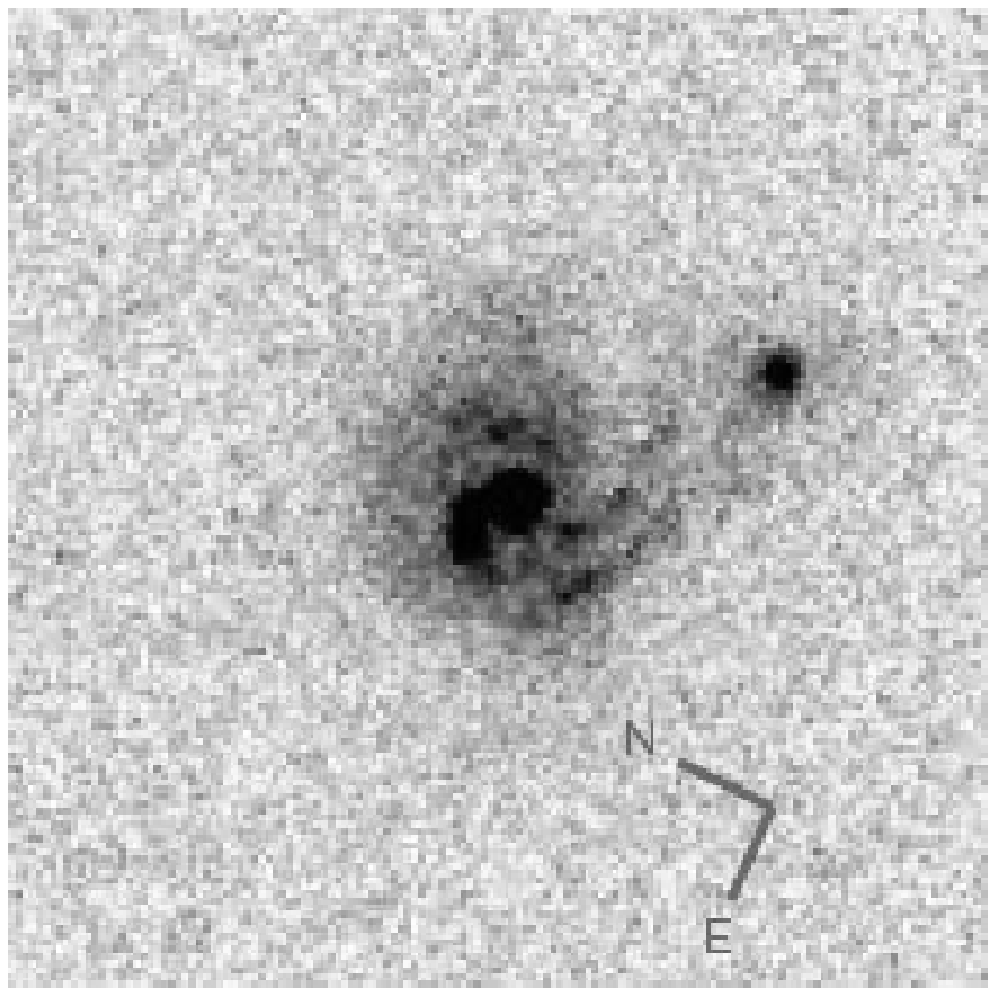}
\hspace*{0.1cm}
\includegraphics[height=2.87cm]{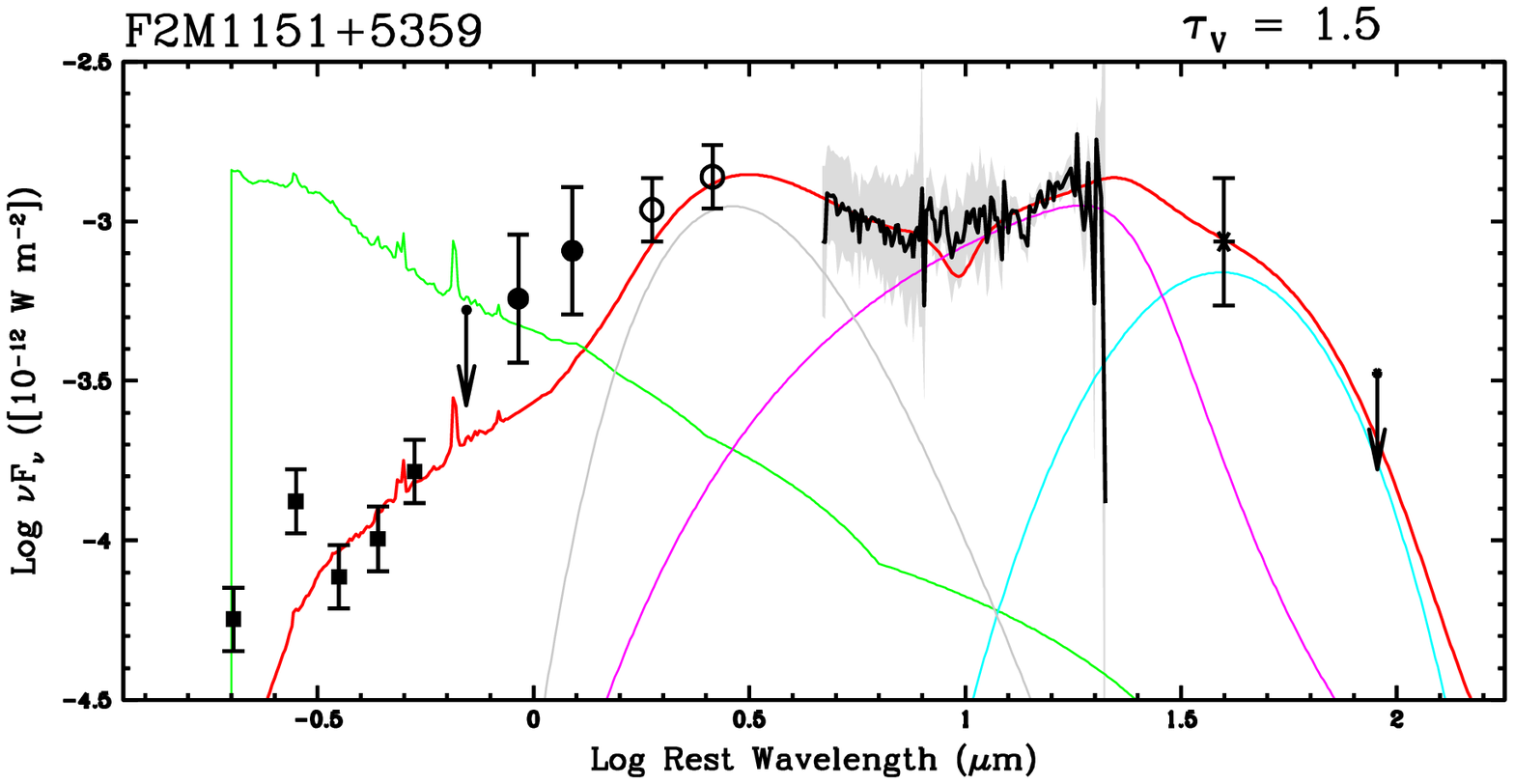}
\includegraphics[width=2.38cm]{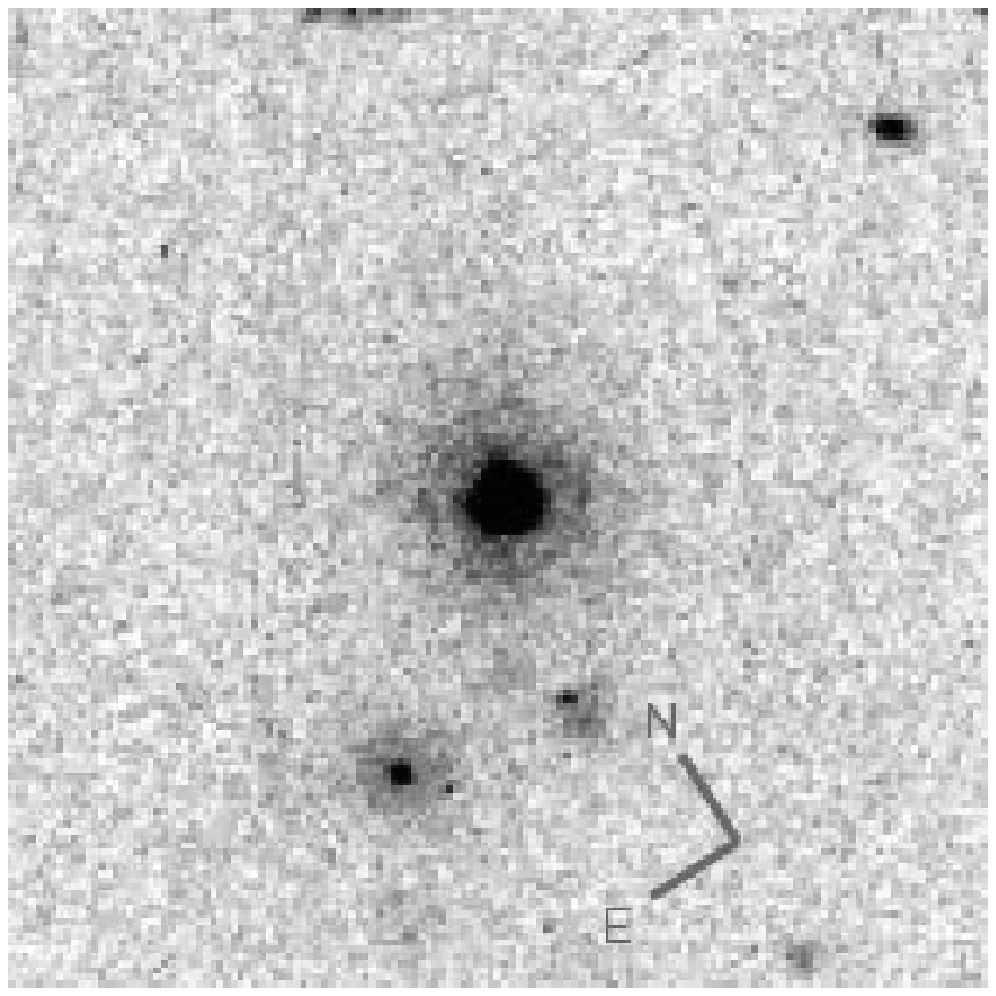}
\includegraphics[height=2.87cm]{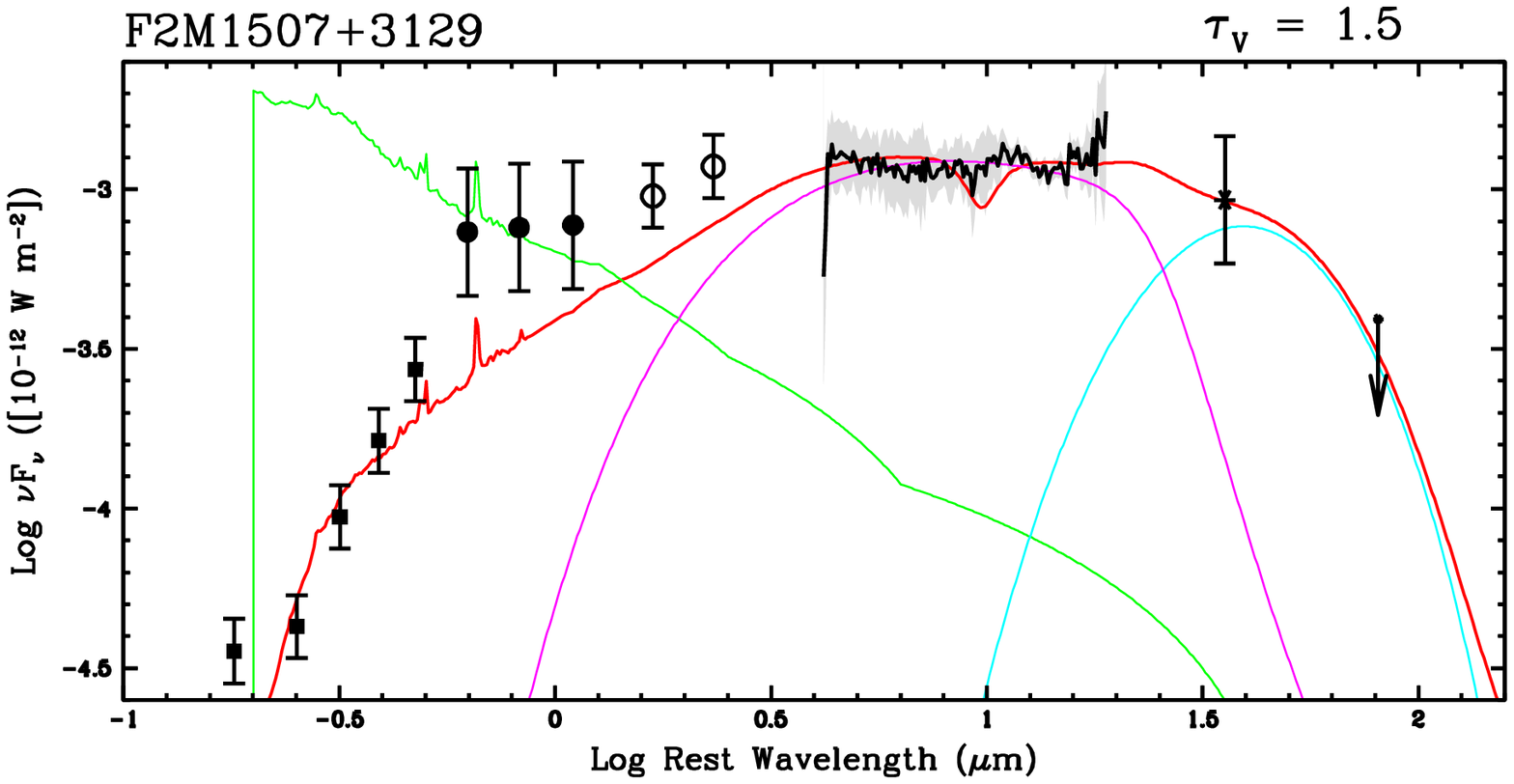}
\includegraphics[width=2.38cm]{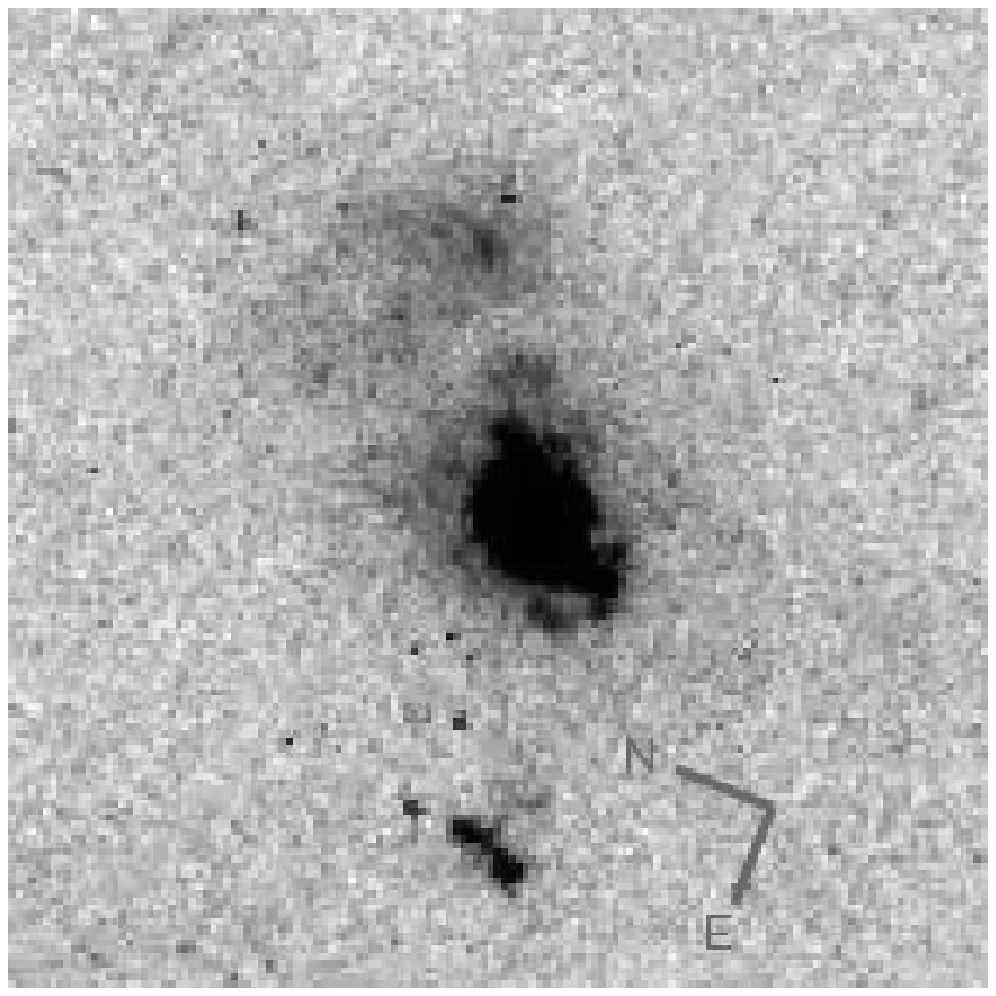}
\hspace*{0.1cm}
\vspace*{0.5cm}
\includegraphics[height=2.87cm]{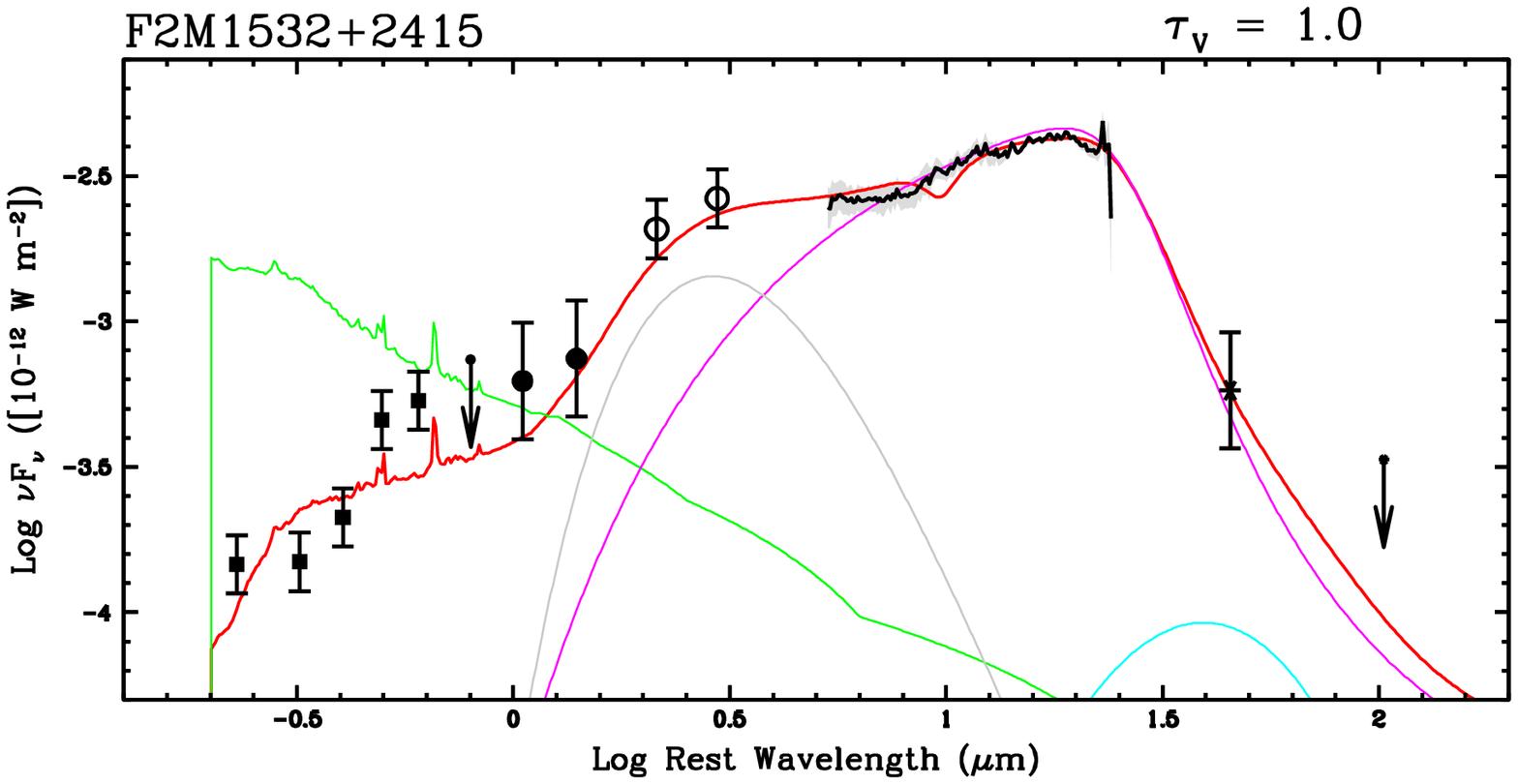}
\includegraphics[width=2.38cm]{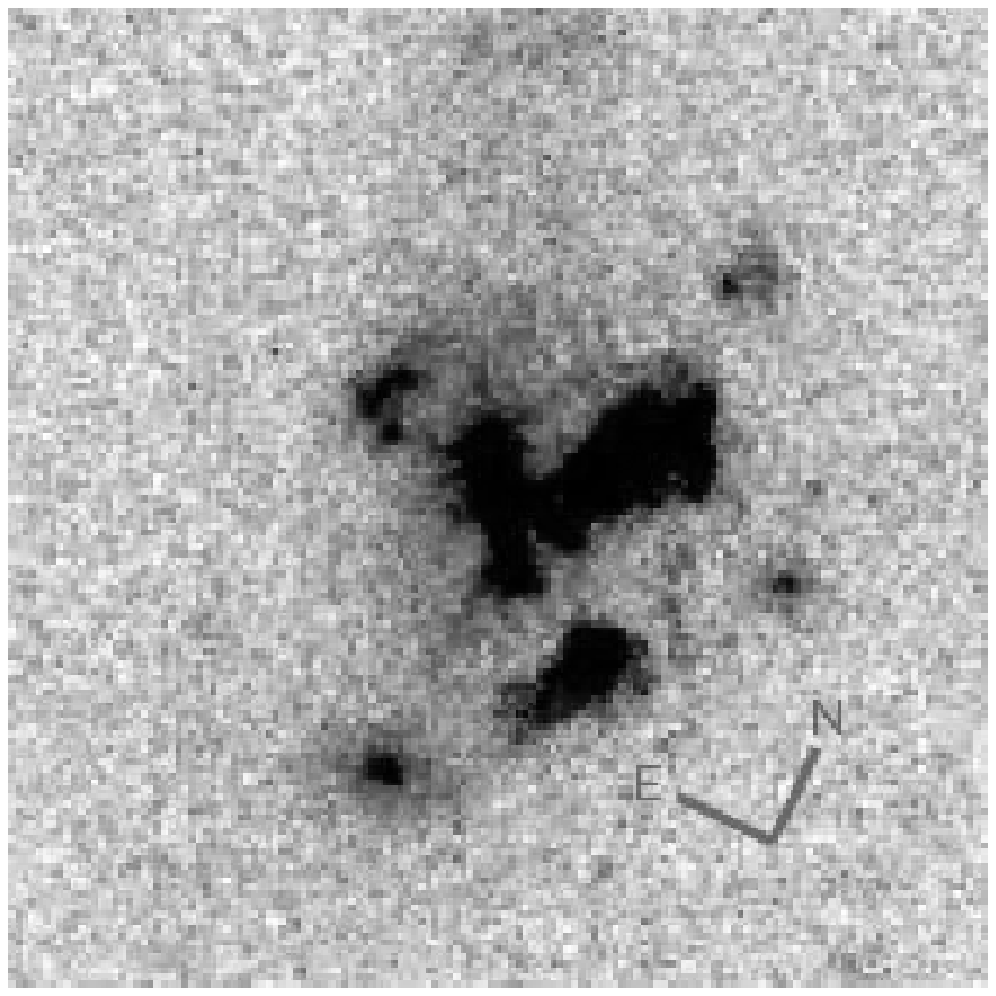}
\includegraphics[height=2.87cm]{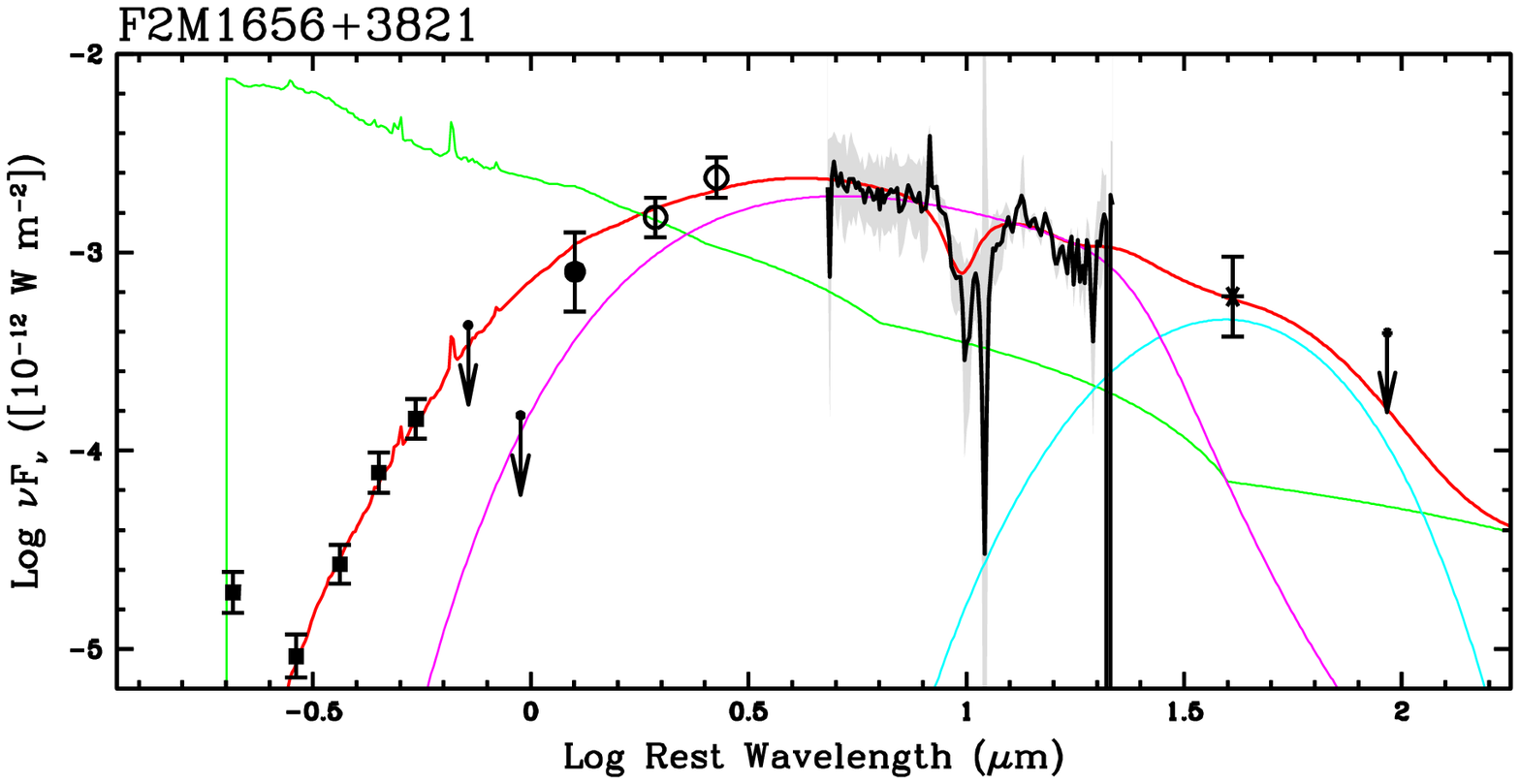}
\includegraphics[width=2.38cm]{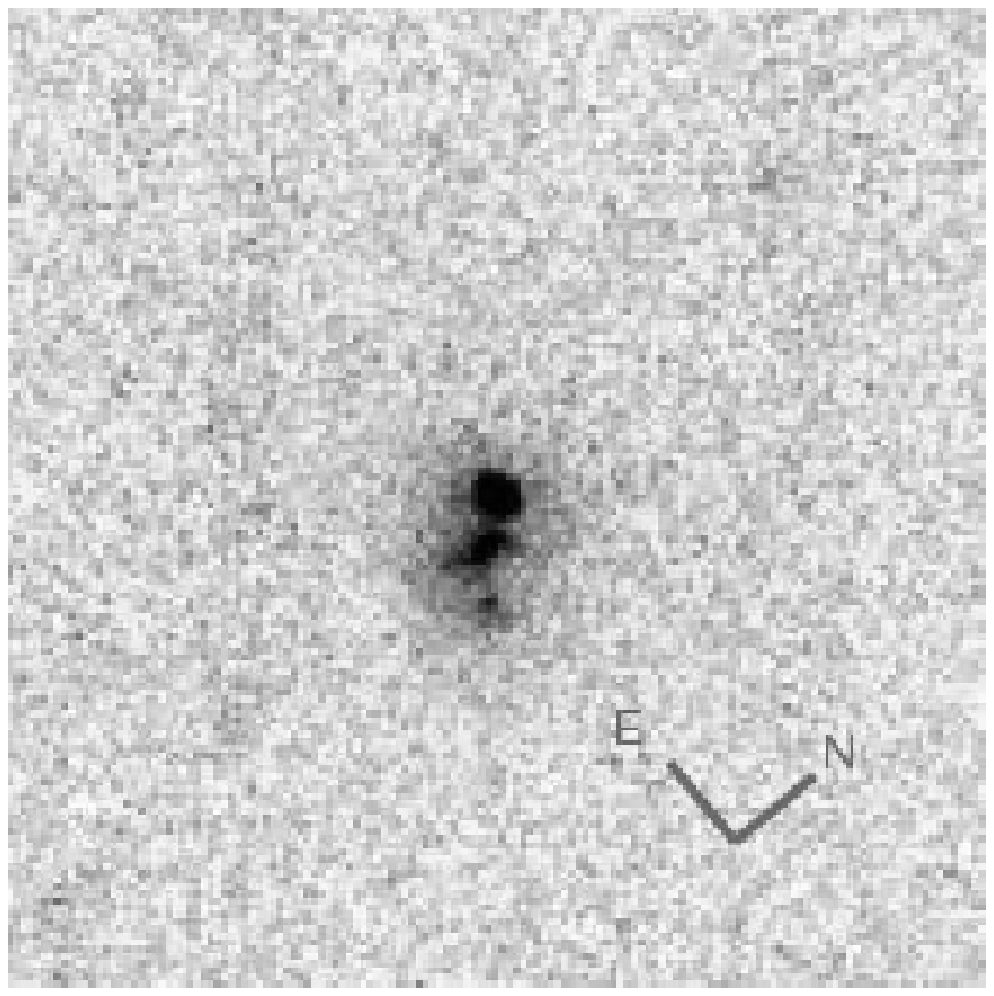}
\caption{Cont.}
\end{center}
\end{figure*}

To model the SEDs we used a phenomenological model based on that of 
\cite{sajina06} due to the current lack of a good physical understanding of 
the nature and geometry of the dust emitting regions. Many of the parameters 
thus do not have a direct physical interpretation. Nevertheless, we 
considered the effort worthwhile as it allows an empirical breakdown of the 
contributions of starlight, hot dust emission from the AGN, and cooler dust 
emission from star formation. The model fits are shown in Figure 
\ref{comparison}. Full details of the model components as applied to quasars 
are given in \cite{hiner09}, but are summarized below. 

(1) The composite quasar spectrum. 
As the composite of \cite{sdssq01} suffers from noticeable host galaxy 
contamination at the long wavelength end, we constructed a new composite by 
subtracting the continuum from the composite of \cite{sdssq01}, and adding 
the residual emission line composite to a new continuum constructed using 
line-free optical/near-infrared SED points from the composite of 
\cite{sdssq}. In Figure \ref{comparison} this is the green line.

(2) A power-law component for the mid-infrared emission, with an exponential 
cutoff at short wavelengths to represent dust sublimation, and a Fermi 
function cutoff at long wavelengths. The functional form used was: 
\begin{equation}
L_{\rm AGN}=\frac{L^{0}_{\rm AGN}\, \nu^{(1-\alpha)}\, 
e^{h \nu / k_B T_{\rm sub}}}{e^{(\nu - \nu_{\rm hcut})/w}+1}
\end{equation}
where the normalization, $L^0_{\rm AGN}$, power-law index $\alpha$ and 
$T_{\rm sub}$ (a proxy for the sublimation temperature) are allowed to vary, 
and $\nu_{\rm hcut}$ and $w$ were fixed at $0.11\times 10^{14}$Hz and 
$0.017\times 10^{14}$Hz, respectively ($h$ and $k_{\rm B}$ are the Planck and 
Boltzman constants, respectively). This component is represented by the 
magenta line in Figure \ref{comparison}. After addition, components (1) and 
(2) were reddened by the Galactic Center extinction curve of \cite{chiar06} 
to fit any silicate absorption feature. 

(3) A warm dust component to represent the small grain emission from 
H{\sc ii} regions, represented by a power-law with fixed high and low 
frequency cutoffs:
\begin{equation}
L_{\rm SG} = \frac{L^{0}_{\rm SG}\, \nu^{(1-\gamma)}\, 
e^{-\nu/\nu_{\rm sgl}}}{e^{-\nu_{\rm sgh}/\nu}}
\end{equation}
this component is poorly constrained in most of the fits, so $\gamma$ was 
fixed at a typical value of two (e.g. \cite{sajina06}). The high and low 
frequency cutoffs, $\nu_{\rm sgl}$ and $\nu_{\rm sgh}$ were set to 
$0.062\times 10^{14}$Hz and $0.17\times 10^{14}$Hz, respectively. Figure 
\ref{comparison} shows this component in light blue.

(4) a thermal greybody model for the far-infrared emission:
\begin{equation}
L_{\rm FIR} = \frac{L^0_{\rm FIR}\, \nu^{3+\beta}}{e^{h\nu/k_{\rm B}T}-1}
\end{equation}
For all of the objects, we fixed the temperature of this component, $T$, to 
40K, because of the uncertainty in our 160\micron~photometry. $\beta$ was 
fixed at 2.0 as above \citep{dunne01}. In Figure \ref{comparison} this is 
represented with a dark blue line. 

(5) Many quasars objects also required an extra near-infrared component to 
match the SEDs. This ``very hot'' component was modeled as a 1000K black body 
(equation 4 with $T=1000$). The physical origin of this component is unclear, 
but it is most likely due to hot dust close to the sublimation radius 
\citep[e.g.,][]{glikman06,quest2,mt11}. This last component can be seen in 
grey color in Figure \ref{comparison}.

In addition, we tried to add a PAH model as described in \cite{lacy07b}, 
however PAH emission was relatively minimal and did not improve our fits. We 
therefore excluded this component in the fitting process. The contribution 
from starlight is so small compared to the quasar and star formation 
components at all wavelengths except bluewards of $\approx$5000\AA, so we 
decided to ignore the stellar contribution in our fitting process.

\begin{deluxetable*}{lccccccc}
\tabletypesize\scriptsize
\tablecaption{Parameter results from multiwavelength fitting \label{fitting}}
\tablewidth{0pt}
\tablehead{
\colhead{Source} & \colhead{QSO contrib-} & \colhead{Starburst (FIR)} & 
\colhead{QSO / SB} & \colhead{E(B-V)\tablenotemark{a}} & 
\colhead{E(B-V)\tablenotemark{b}} & \colhead{$S_{\rm Sil}$} &
\colhead{$D$\tablenotemark{c}} \\
\colhead{} & \colhead{ution (intrinsic)} & \colhead{contribution} & 
\colhead{Ratio} & \colhead{(spectrum)} & \colhead{(SED fit)} & \colhead{} & 
\colhead{coefficient} \\ 
\colhead{} & \colhead{(log $L_{QSO}/L_{\odot}$)} & 
\colhead{(log $L_{FIR}/L_{\odot}$)} & \colhead{} & \colhead{} & \colhead{} & 
\colhead{}& \colhead{}} 
\startdata
F2M0729$+$3336 & 12.83 & $<$11.47 &$>$ 22.91 & 0.83$\pm$0.22 & 0.605 & 
-0.43$\pm$ 0.09& 1.92 \\
F2M0825$+$4716 & 13.10 & 12.35 &  5.62 & 0.52$\pm$0.10       & 0.806 & 
-0.50$\pm$ 0.03&  1.29 \\
F2M0830$+$3759 & 12.03 & 11.35 &  4.79 & 0.80$\pm$0.15       & 0.478 &  
 0.00$\pm$ 0.06& 1.39 \\
F2M0834$+$3506 & 12.04 & 11.42 &  4.17 & 0.58$\pm$0.05       & 0.629 & 
 0.14$\pm$0.03& 0.95 \\
F2M0841$+$3604 & 12.12 & 12.03 &  1.23 & 1.34$\pm$0.11       & 0.770 & 
-0.76$\pm$ 0.15& 1.73 \\
F2M0915$-$2418 & 13.37 & 12.69 &  4.79 & 0.36$\pm$0.12       & 0.354$^*$ & 
-0.05$\pm$ 0.05& 1.29 \\
F2M1012$+$2825 & 12.24 & $<$11.88 & $>$2.29 & 0.82$\pm$0.10  & 0.647 & 
 0.03$\pm$ 0.10& 1.13 \\
F2M1113$+$1244 & 13.14 & 12.43 &  5.13 & 1.41$\pm$0.11       & 0.294 & 
-0.31$\pm$ 0.01& 1.24 \\
F2M1118$-$0033 & 12.71 & 12.29 &  2.63 & 0.85$\pm$0.11       & 0.512 & 
-0.15$\pm$ 0.04& 1.40 \\
F2M1151$+$5359 & 12.15 & 11.77 &  2.40 & 0.42$\pm$0.08       & 0.532$^*$ & 
-0.01$\pm$ 0.07& 0.95 \\
F2M1507$+$3129 & 12.57 & 12.05 &  3.31 & 0.74$\pm$0.09       & 0.532$^*$ & 
-0.01$\pm$ 0.07& 1.16 \\
F2M1532$+$2415 & 12.47 & 10.67 & 63.10 & 0.90$\pm$0.53       & 0.354$^*$ & 
-0.09$\pm$ 0.02& 1.50 \\
F2M1656$+$3821 & 12.44 & 10.83 & 40.74 & 0.88$\pm$0.16       & 1.169 & 
-1.31$\pm$ 0.01& 1.34 \\
\enddata
\tablenotetext{a}{Derived via spectral fitting of a quasar composite reddened 
by the SMC law, either from \cite{redqso-hst} or \cite{f2ms}; the latter 
involves much more careful fitting, masking emission line regions.}
\tablenotetext{b}{Derived from the SED fitting and modeling as described in 
Section \ref{model}, the extinction ($\tau_V$) is derived from Silicate 
absorption, except for objects marked with a star. Note the general agreement 
with the Silicate depth derived via the method of \cite{spoon07}.}
\tablenotetext{c}{$D=G/C$, used as disturbance parameter. Values of G and C 
are from \cite{redqso-hst}, see that publication for details on how the 
values were obtained. D values close to 1.0 represent morphologically 
undisturbed systems. The higher the D value, the more disturbed the system is.}
\end{deluxetable*}

Most of our quasars are well-fit by this model, with relatively small 
$\chi^2/dof$ values, but in some cases the fit at the near-IR regions of the 
spectrum described by 2MASS magnitudes provides is poor. This is most likely 
due to the host galaxy emission dominating the optical and/or near-infrared 
passbands. 

In the cases where the model fit the broadband SED well, the reddening values 
derived from the Silicate absorption feature were in general agreement to the 
values of $E(B-V)$ derived from the SMC continuum fit to the optical spectrum 
\citep{f2m07,f2ms}, confirming cold dust as the absorbing and reddening 
mechanism in these systems (see Table \ref{fitting}). However, in four cases 
(F2M0915+2418, F2M1151+5359, F2M1507+3129 and F2M1532+2415) there was no 
Silicate in absorption to account for the reddening seen in the optical and 
near-infrared. In those cases, we artificially fixed the extinction $\tau_V$ 
that would've been derived from the fit from the IRS spectrum manually. These 
objects have their modeled extinctions marked with a star in Table 
\ref{fitting} and their fixed extinctions labeld in Figure \ref{comparison}.

We then derived the obscured quasar luminosities by integrating over the 
quasar components (1)+(2); the intrinsic luminosities quoted in Table 
\ref{fitting} are then derived by accounting for the extinction obtained from 
the Silicate absorption fit. Since the AGN contribution is well constrained 
by the IRS spectrum and optical/near-IR photometry points, the uncertainties 
are overall small ($\lesssim 0.2$ dex) and only pertain to the accuracy of the 
Silicate absorption feature. The FIR/Starburst contribution was taken solely 
from the integration of the two dust bumps (light and dark blue; (3)+(4)), 
not from any component in the optical regime. Since that FIR contribution is 
largely determined by the 70 and 160 \micron$\;$ photometry points alone, it 
has the largest uncertainties in the contribution ($\simeq 0.5$ dex). The 
Starburst contribution ranges from $\log(L_{\rm FIR}/L_\odot)$ = 10.67 -- 
12.67, placing the host galaxies of red quasars roughly in the LIRG regime. 
For the QSO contribution, we find them to have bolometric luminosities in the 
range of $\log(L_{\rm QSO}/L_\odot) =$ 12.03 -- 13.37. These are very large 
values, placing these quasars among the brightest in the Universe in the IR 
furthermore strengthening our previous conclusion that we are only probing 
the tip of the quasar luminosity iceberg \citep{eilat04,f2m07,f2ms}.

\subsection{Black hole masses and accretion rates}

We estimated black hole masses ($M_{\rm BH}$) for the twelve of our thirteen 
quasars for which broad lines were visible in either the optical or 
near-infrared spectra (Table \ref{bhmasses}). H$\beta$ was our preferred line 
as it is relatively isolated, and most quasar black hole mass estimates are 
calibrated on it, but in most cases the broad H$\beta$ line was too heavily 
extinguished, so a longer wavelength line such as H$\alpha$ or Pa$\beta$ was 
used (Pa$\beta$ was generally preferred as it is free of contaminating narrow 
lines). This assumed that the H$\beta$, H$\alpha$ and Pa$\beta$ lines have 
the same velocity widths, but was indeed the case (within the errors) for the 
objects in which we could measure two or more broad line widths 
(F2M0825+4716, H$\alpha$ and Pa$\beta$; F2M0834+3506, H$\beta$ and H$\alpha$; 
F2M0915+2418, H$\alpha$ and Pa$\beta$; F2M1113+1244, H$\beta$, H$\alpha$ and 
Pa$\beta$). 

\begin{center}
\begin{deluxetable*}{lcccccccc}
\tabletypesize\scriptsize
\tablecaption{Black hole mass estimates and Eddington rates \label{bhmasses}}
\tablewidth{0pt}
\tablehead{
\colhead{Object} & \colhead{Redshift} &  \colhead{Line} & \colhead{FWHM} &
\colhead{${\rm log}$} & \colhead{${\rm log}$} & 
\colhead{${\rm log}$} & 
\colhead{${\rm log_{10}}$} & 
\colhead{${\rm log_{10}}$} \\
\colhead{} & \colhead{} & \colhead{} & \colhead{(km s$^{-1}$)} & 
\colhead{$(\nu L_{15})$} & \colhead{($M_{\rm BH}/M_{\odot}$)} & 
\colhead{$(L/L_{\rm Edd})$} & \colhead{$(L_B/L_{\odot})$} & 
\colhead{$(L_B/L_{\odot})$}\\
\colhead{} & \colhead{} & \colhead{} & \colhead{} & 
\colhead{(erg s$^{-1}$)\tablenotemark{a}} & \colhead{} & \colhead{} & \colhead{(fit)\tablenotemark{b}} & 
\colhead{(monotonic)\tablenotemark{c}}}
\startdata
F2M0729$+$3336 & 0.957 & H$\beta$  & 2866$\pm$ 200  & 45.8 & 8.7$\pm$0.1 & 
0.1$\pm$0.1 & 11.06 & 10.99 \\
F2M0825$+$4716 & 0.803 & Pa$\beta$ & 2664$\pm$ 100  & 46.1 & 8.8$\pm$0.1 & 
0.3$\pm$0.1&10.30&10.90\\
F2M0830$+$3759 & 0.414 & H$\beta$  & 3741$\pm$ 200  & 45.3 & 8.6$\pm$0.1 &
-0.4$\pm$0.1&10.97&11.01\\
F2M0834$+$3506 & 0.470 & H$\alpha$ & 10500$\pm$2000 & 45.2 & 9.5$\pm$0.1 &
-1.3$\pm$0.1&10.69&10.65\\
F2M0841$+$3604 & 0.553 & H$\alpha$ & 3380$\pm$ 500  & 45.2 & 8.5$\pm$0.1 &
-0.4$\pm$0.1&10.53&10.73\\
F2M0915$+$2418 & 0.842 & Pa$\beta$ & 4186$\pm$800   & 46.5 & 9.4$\pm$0.2 & 
0.1$\pm$0.1&10.58&10.66\\
F2M1012$+$2825 & 0.937 & H$\alpha$ & 8050$\pm$ 1000 & 45.3 & 9.3$\pm$0.1 &
-1.0$\pm$0.1&10.62&10.79\\
F2M1113$+$1244 & 0.681 & H$\beta$  & 2276$\pm$ 300  & 46.1 & 8.6$\pm$0.1 & 
0.4$\pm$0.1&10.91&10.91\\
F2M1118$-$0033 & 0.686 & Pa$\beta$ & -              & 45.7 & $<$8.7$^d$      &
-&11.01&11.12\\
F2M1151$+$5359 & 0.780 & H$\alpha$,Pa$\beta$ & 5700$\pm$500&45.4&9.0$\pm$0.1 &
-0.7$\pm$0.1&10.48&10.88\\
F2M1507$+$3129 & 0.988 & H$\alpha$ & 12800$\pm$2000 & 45.7 & 9.9$\pm$0.1 &
-1.3$\pm 0.1$&10.60&10.96\\
F2M1532$+$2415 & 0.562 & Pa$\beta$ & 11700$\pm$1000 & 45.6 & 9.8$\pm$0.1 & 
-1.2$\pm$0.1&10.95&11.09\\
F2M1656$+$3821 & 0.732 & Pa$\beta$ & 3064$\pm$300   & 45.4 & 8.5$\pm$0.1 &
-0.2$\pm$0.1&10.54&10.67\\
\enddata
\tablenotetext{a}{monochromatic luminosity at 15$\mu$m (rest).}
\tablenotetext{b}{rest-frame $B$-band luminosity of the central component of 
the host galaxy as given in table 6 of \cite{redqso-hst} (``Total'' host 
magnitude in table 6).}
\tablenotetext{c}{rest-frame $B$-band luminosity of the extended host galaxy 
(all components) (``mono'' luminosity in table 6). The $B$-band luminosities 
were estimated from the observed $I$-band fluxes assuming a stellar SED flat 
in $F_{\lambda}$ to perform the small k-corrections required to map observed 
$I$-band to rest-frame $B$-band.}
\tablenotetext{d}{limit on the black hole mass assuming Eddington-limited 
accretion.}
\end{deluxetable*}
\end{center}

Black hole mass estimates were then calculated using the broad-line widths 
and quasar luminosity assuming a broad line region radius, $R_{\rm BLR}$:
\begin{equation}
R_{\rm BLR}= K + \alpha{\rm log}(\lambda L_{\lambda}(5100A)),
\end{equation}
\citep{bentz09} with $K=-21.3$ and $\alpha=0.519$ and estimating the 
intrinsic luminosity at 5500\AA$\;$,$\lambda L_{\lambda}(5100A)$ by measuring 
the quasar luminosity at a rest-frame wavelength of 15$\mu$m (where 
extinction should be negligible) and applying the ratio of the bolometric 
corrections at 5100$\AA \;$ (12) to the correction at rest-frame 15$\mu$m 
(9.0) from \citep{sdssq}. The black hole mass estimate, $M_{\rm BH}$ is then 
given by:
\begin{equation}
M_{\rm BH} = f\frac{\sigma^2}{GR_{\rm BLR}}
\end{equation}
where $\sigma$ is the velocity dispersion of the broad line, $G$ is the 
gravitational constant, and we assume a geometric correction factor of 
$f=5.5$. The Eddington luminosity is: 
\begin{equation}
L_{\rm Edd} = 1.3\times 10^{31}\left(\frac{M_{\rm BH}}{M_{\odot}}\right) 
{\rm W},
\end{equation}
and Eddington ratios are estimated by dividing the intrinsic luminosity of 
the quasar from the fit by the Eddington luminosity. 

\section{Results}

We first place our red quasars into the context of other luminous infrared 
galaxies using the \cite{spoon07} diagnostic diagram (Figure 
\ref{diagnostics}). This plot of silicate absorption depth versus PAH 
equivalent width separates the known classes of infrared-luminous galaxies 
into two. One sequence extends from region 1C (comprised mostly of starburst 
dominated objects) into 1A (comprised mostly of IR-bright quasars) as 
continuum replaces PAH emission in objects with little silicate emission or 
absorption. We plot this sequence with cyan double arrows in Figure 
\ref{diagnostics}. The second sequence extends from 1C through 2B through 3A 
in which objects again become continuum dominated, but with high silicate 
absorption depths (the ``warm ULIRG'' population), shown with magenta double 
arrows in Figure \ref{diagnostics}. 

Normal quasars are typically found in region 1A, as are the majority of our 
dust-reddened objects, but the fact that we see one object firmly and one 
barely in the the sparsely populated quadrant 2A of the diagram is 
noteworthy. Such objects are rare in most compilations of galaxies and AGN 
and are believed to be part of a transitioning population, confirming that we 
have indeed located an interesting class of object, perhaps intermediate 
between the heavily absorbed warm ULIRGs and the normal quasar population. 
Also, it is worth mentioning that none of the sources show silicate in 
emission, which is unusual for these very IR-bright quasars; it's possible 
that in some of the objects (like the four we had to fix the Silicate 
absorption manually in the SED fitting) the silicate absortion is missing 
because it is being filled with an emission bump.

The results of the fitting (Figure \ref{comparison} and Table \ref{fitting}) 
show that the AGN component dominates the bolometric luminosity in all cases, 
with most of our objects having measurements or limits on their far-infrared 
luminosities corresponding to star formation rates of 
$\sim 10-300M_{\odot}{\rm yr^{-1}}$ using the \cite{kennicutt} conversion 
between far-infrared luminosity and star formation rate. Such star formation 
rates indicate luminous starbursts, but not at the level at which they are 
adding significantly to the stellar mass of the host galaxies.

Next we consider the accretion luminosities of our objects. Our objects span 
a wide range of luminosities relative to the Eddington Limit ($L_{\rm Edd}$),
including many objects with high luminosities when compared to typical quasar 
Eddington Ratios $\sim 0.1$ \citep[e.g.,][]{steinhardt10}. This could be an 
artifact of partial obscuration,  for example, if the inner broad line 
region, where the linewidths are greatest, remains somewhat obscured. 
However, where we are able to measure the widths of multiple broad lines in 
the same object, the linewidths are very similar irrespective of wavelength, 
making this latter possibility unlikely. 

\begin{figure*}
\plotone{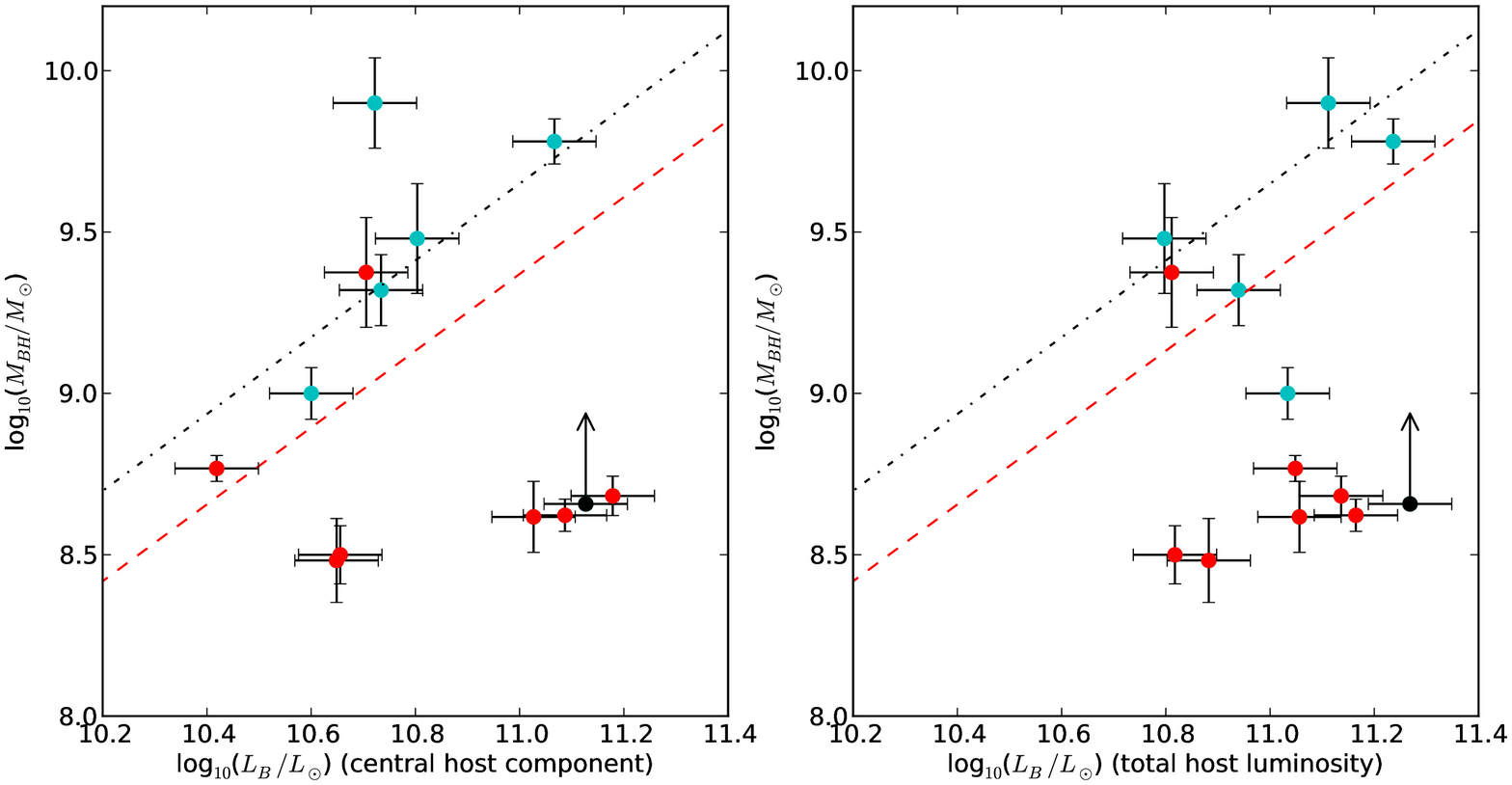}
\caption{Black hole mass versus host luminosity for our red quasars. 
{\em Left:} using the model fits to the brightest host galaxy component of 
\cite{redqso-hst}, {\em right:} using the total host luminosity, including 
the flux in disturbed, outlying components not well-fit by the model. Cyan 
symbols denote objects accreting at $<$30\% of the Eddington rate, red points 
objects with accretion rates $>$30\% Eddington, the object with a limit only
on its black hole mass (assuming Eddington-limited accretion), 
F2M1118-0033, is shown as a black dot. 
The red dashed line is the 
local black hole mass -- bulge luminosity relation from \cite{vdM98}, the 
black dashed line is the relationship at $z\sim 0.7$ 
\citep[interpolated from][]{merloni10}. Most of our high accreting object lie 
below the black hole mass -- bulge luminosity relation.\label{mbhlstar}}
\end{figure*}

Figure \ref{mbhlstar} displays the black hole mass of our quasars versus 
their host luminosity, left for the central bulge component and right for the 
total host light. The central bulge luminosity was derived by fitting the 
brightest host galaxy components, while the total luminosity component 
included the flux in disturbed, outlying components, both after subtraction 
of the quasar nucleus (see \cite{redqso-hst} for an in depth discussion on 
the AGN -- host galaxy fitting). Objects with high accretion rates (above 0.3 
Eddington) are plotted in red, objects with low rates in blue. The Figure 
shows that there is very little correlation between the black hole masses and 
the corresponding galaxy luminosities, either of the fitted central galaxy 
component or to the total host galaxy light. However, none of our black hole 
masses exceed the black hole mass -- bulge luminosity relation 
\citep{marconihunt} by a large factor, also when corrected for evolution 
according to \cite{bennert10}, even when only the central component of a 
merging system (defined as the one containing the quasar nucleus) is 
considered. The fact that we find a significant fraction of our objects with 
high accretion rates below the black hole mass -- bulge luminosity relation 
leads us to speculate that they are young, just ignited black holes. Their 
growth has started after the host galaxy growth via a starburst event, yet 
since they are accreting at such high rates they will move onto the 
``classical'' relation within $\sim$10$^7$ years.

\begin{figure*}
\plotone{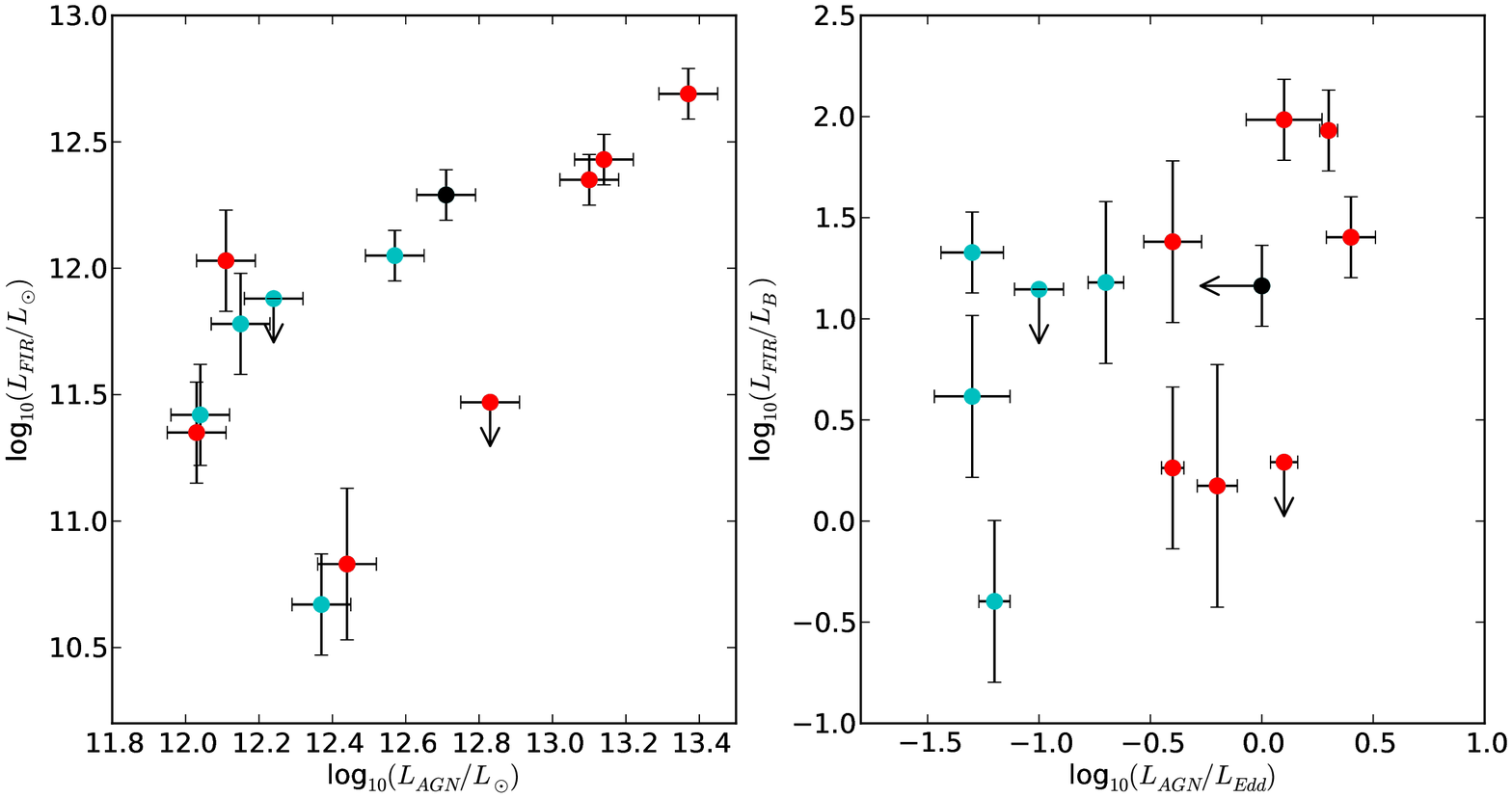}
\caption{The log of the ratio of far-infrared luminosity to host luminosity 
in $B$-band (as a proxy for the specific star formation rate) is plotted 
against the accretion rate relative to the Eddington limit. Symbols are 
as in Figure 4. While the most 
luminous quasars also show the highest luminosity in the FIR, there is no 
correlation between the inferred star formation and accretion rate of the 
quasar.\label{ssfrledd}} 
\end{figure*}

In Figure \ref{ssfrledd} we plot both the total FIR-luminosity against the 
AGN luminosity, and the ratio of accretion rate to the Eddington limit of the 
quasar against the ratio of the far-infrared luminosity to the $B$-band 
luminosity of the host galaxy (a proxy for the specific star formation rate). 
There is a hint of a correlation between the total FIR luminosity and AGN 
luminosity \citep[as found for nearby quasars,][]{quest2}, though it is not 
tight. There seems to be no relation between the mass-normalized quantities 
of specific star formation rate and accretion rate relative to the Eddington 
Limit.

Finally, in Figure \ref{morphcorr}, we examine the relationship between the 
host galaxy morphology and the star formation rate and AGN accretion rate. In 
order to quantify the degree of disturbance of the host we use the ratio of 
Gini Coefficient (G) to Concentration Index (C). We showed in 
\cite{redqso-hst} that the combination of these two quantities gave a robust 
indication of the degree of disturbance of the host galaxy. We therefore 
define a ``disturbance parameter'', $D=G/C$. This quantity is approximately 
unity for normal galaxies \citep{cas}, but is significantly greater than 
unity for highly disturbed systems. There is a weak (though not statistically
significant) trend for high accretion rate objects to have more disturbed 
morphologies, though there is no apparent trend with specific star formation
rate.

\section{Discussion}

Our aim in this study was to determine whether the infrared properties of 
these dust-reddened objects were consistent with the merger-driven 
co-evolutionary model for the growth of SMBH and their host galaxies. In 
particular, whether these objects are an intermediate stage between a 
merger-induced starburst and a ``normal'' quasar with an unreddened, blue 
continuum, with AGN feedback having stopped further star formation in the 
host. Our results seem to be consistent with this overall picture, but we see 
considerable variation in properties such as silicate absorption depth, star 
formation rate and AGN accretion rate on an object-by-object basis.

\begin{figure*}
\plotone{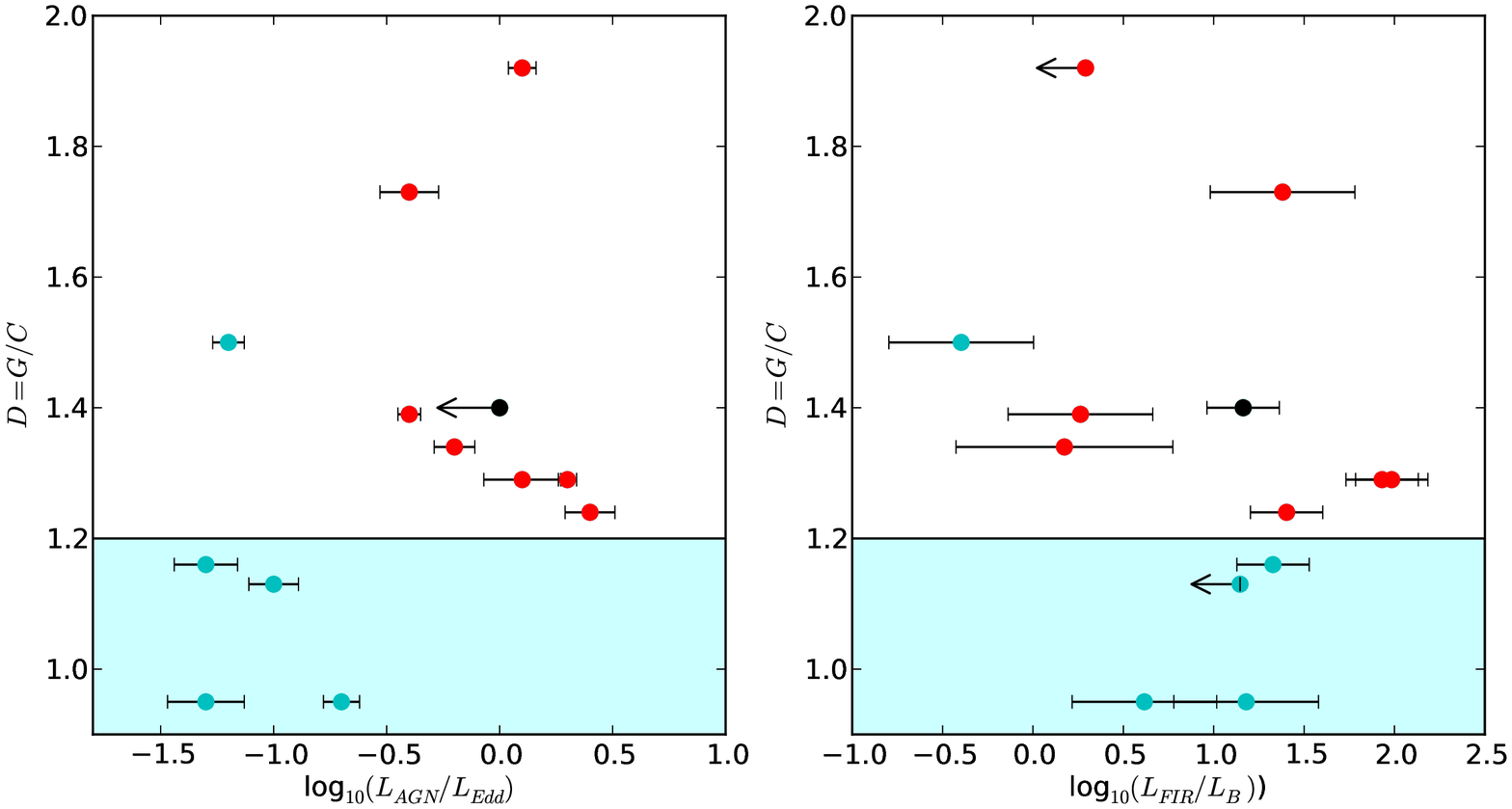}
\caption{{\em Right}: Host galaxy morphological distortion, parameterized 
as $D$, the ratio of the Gini Coefficient (G) to the Concentration Index (C) 
against Eddington ratio. Normal galaxies have a mean 
$D\approx 1.2$ (Abraham et al.\ 2003), objects with values below this 
(shaded in the plot) are relatively undisturbed, those with higher values
have increasing amounts of disturbance.
Symbol colors as for Figure \ref{mbhlstar}. {\em Left:} 
Morphological distortion versus the log of the ratio of far-infrared 
luminosity to host luminosity in $B$-band, an indicator for star formation. 
While there could be a hint of a correlation between the amount of host 
disturbance and the accretion rate, there is no correlation found for star 
formation versus the host disturbance.\label{morphcorr}}
\end{figure*}

Much of the object to object variation is due to the heterogeneity of our 
sample; some have high reddenings, morphological disturbances, some do not. 
While they all are luminous quasars, they do span a large range in 
luminosities, too. In the overall picture, however, trends are evident. If an 
object is morphologically disturbed, it is much more likely to have a high 
Eddington ratio and to show Silicate in absorption (be reddened by cold 
dust). Similarly, the two quasars which did not show any signs of merger 
interaction in \cite{redqso-hst} also do not display high Eddington ratios or 
Silicate in absorption and therefore should not be included characterized as 
``young quasars''. Not all dust-obscured quasars fit in the co-evolutionary 
scenario, but statistically, it is the most likely explanation and therefore 
our conclusions are to be taken as such.

All these objects have mid-infrared (AGN) luminosities in excess of their 
far-infrared (presumed starburst) luminosities. This dominance of the AGN 
suggests that AGN feedback processes, if they are indeed important, will have 
started to have their impact on star formation in the host galaxy as we would 
have expected the starburst component to have dominated otherwise. We do not 
see any obvious quasar feedback ``smoking guns''; broad absorption line 
features from AGN thermal winds, for example, are not present in any of our 
objects which are of high enough redshift to see \ion{Mg}{2} absorption 
($z\stackrel{>}{_{\sim}}$0.7). However, experience from \cite{f2ms}, has 
shown that for the red quasars selected using our criteria the spectrum below 
$\approx$ 5000 \AA~is so extinguished that even deep absorption features are 
difficult to detect in objects with $z\stackrel{<}{_{\sim}}$0.9. Only three 
of our objects (F2M0729+3336, F2M1012+2825 and F2M1507+3129) have appropriate 
redshifts, and F2M0729+3336 is too noisy to detect absorption features. 
Nevertheless, an excess of low-ionization, broad absorption line quasars has 
been found in our full sample of red quasars, \cite{f2ms}, and a tentative 
relationship between wind properties and star formation seen has been seen by 
\cite{farrah11} in samples of reddened quasars. 

Blueshifted wings to [\ion{O}{3}]5007\AA, with velocities relative to the line 
peak of up to $\sim$1000 km s$^{-1}$ are seen in five of our thirteen objects 
(F2M0729+3336, F2M0825+4716, F2M0915+2418, F2M1118-0033, F2M1151+5359). Of 
the four of these for which we can estimate accretion rates, three have 
accretion rates just exceeding the Eddington limit, possibly indicating a link 
with extremely high accretion rates, but the star formation rates in the 
hosts vary widely. We also lack any information on the spatial extent of 
these outflows, which may be confined to the nuclear regions. 
Blueshifted [\ion{O}{3}] line components are common in compact radio sources: 
\cite{htm08} propose that these components are due to outflows of gas driven 
by the radio jets. The radio sources in our quasars are however significantly 
less luminous than those in the radio galaxies studied by \cite{htm08}. 
Radio-quiet type-2 luminous quasars at low redshift also tend to show 
outflows in [\ion{O}{3}]5007\AA~emission \citep{greene11a}, suggesting that 
these are not confined to radio sources, though the most spectacular object 
in the \cite{greene11a} sample, J1356+1026, is a relatively strong radio 
source \citep{greene11b}. Clearly the role of radio jets in powering these 
outflows needs to be established.

In agreement with the evolutionary model, we see a large fraction of objects 
with high black hole accretion rates, consistent with the idea that quasar 
accretion rates are high in the early phase of growth. The objects with the 
highest accretion rates also tend to fall below the local black hole mass - 
bulge luminosity relation by $\sim 0.5$dex; none of our objects with lower 
accretion rates are found there. This offset from the local relation is 
consistent with a picture in which the black hole is yet to grow to its 
equilibrium size following a major merger. It is also broadly consistent with 
numerical simulations of black hole growth in quasar merger scenarios, in 
which the total black mass hole grows by about 0.5 dex with every major 
merger event \citep[e.g.,][figure 10]{li07}. \cite{sarria10} also find that 
their luminous, dust-obscured quasars at $z\sim 2$ fall below the black hole 
mass -- bulge mass relation at that redshift. An interesting exception is 
F2M0915+2418, which has a high accretion rate, but falls close to the black 
hole mass -- bulge mass relation. This object also has one of the highest 
star formation rates in the sample, however, allowing it to evolve along a 
trajectory that will keep it close to the black hole mass -- bulge mass 
relation.

We do, however, have some remaining puzzles, where our data do not fit in 
with the most naive expectations of the evolutionary model. In 
\cite{redqso-hst} we saw a relationship between $D$ and the amount of 
reddening in the host, but in Figure \ref{morphcorr} there seems to be little 
correlation between the degree of disturbance of the host galaxy and the star 
formation rate, except that all high accreting quasars are also disturbed to 
some degree. There may, however, be a very weak trend for objects with high 
accretion rates relative to the Eddington Limit to have high values of $D$, 
see Figure \ref{morphcorr}. Similarly, there is no clear morphological trend 
related to the offset of the quasar from the local galaxy luminosity -- black 
hole mass relation. 

To some extent, the lack of clear trends here may be due to an admixture of 
objects reddened by foreground galaxies. In particular the low accretion rate 
objects may include some objects where the reddening is from an intervening 
galaxy through a chance alignment, perhaps from a galaxy in the same group or 
cluster. F2M1532+2415 is a good example of this, with an apparently 
highly-disturbed, early stage merger morphology, but both a low accretion 
rate and a low star formation rate, and F2M0834+3506 is a very good candidate 
for a low accretion rate broad-line radio galaxy reddened by a foreground 
irregular galaxy.

In summary, our observations are consistent with the simplest evolutionary 
models, where a merger induced starburst triggers a obscured quasar which 
begins life accreting at close to the Eddington rate and later evolves into 
an unobscured quasar accreting at the $\sim 0.1$ of the Eddington rate more 
typically observed in the normal quasar population. We see no obvious 
evidence, however, of feedback affecting star formation in this particular 
sample (though we do see evidence of strong outflows of ionized gas in some 
objects). Nor do we see a clear sequence or progression from a newborn, 
highly obscured, high accretion rate AGN with a very disturbed host into a 
quiescent, almost unreddened object with a lower accretion rate. Our quasars 
were, however, selected within a fairly narrow range of luminosity (due to 
the 2MASS detection requirement and restricted redshift range) and reddening 
(if the reddening were too high, they would be classed as type-2s, but if it 
were too low they would not be picked out as red quasars), so we might not 
expect to be able to see much of an evolutionary sequence. 

These observations are most consistent with a picture in which the majority 
of objects undergo the starburst/merger phase well before we see these 
reddened quasars emerge, though not so long after that evidence for 
morphological disturbance of the hosts has been erased. The evolution in the 
black hole mass -- bulge mass plane is then a shift along the bulge mass axis 
by $\sim +0.5$dex followed $\sim 10^{8}$yr later by a shift along the black 
hole mass axis by a similar amount. In one case, however, (F2M0915+2418) we 
do seem to see the starburst and accretion happening simultaneously. As our 
selection techniques for the dust-reddened quasar population expand, 
particularly as mid-infrared selected samples become large enough to include 
significant numbers of reddened type-1 quasars we expect to be able to expand 
our samples considerably, and trace out more of the evolutionary scenario.

\acknowledgments

We thank the Spitzer Science Center for funding through Spitzer program 40143.

This work is based on observations made with the Spitzer Space 
Telescope which is operated by the Jet Propulsion Laboratory, California 
Institute of Technology under a contract with NASA.

This publication makes use of data products from the Wide-field Infrared 
Survey Explorer, which is a joint project of the University of California, 
Los Angeles, and the Jet Propulsion Laboratory/California Institute of 
Technology, funded by the National Aeronautics and Space Administration.

Funding for the SDSS and SDSS-II has been provided by the Alfred P. Sloan 
Foundation, the Participating Institutions, the National Science Foundation, 
the U.S. Department of Energy, the National Aeronautics and Space 
Administration, the Japanese Monbukagakusho, the Max Planck Society, and the 
Higher Education Funding Council for England. The SDSS Web Site is 
http://www.sdss.org/.

This work was partly performed under the auspices of the US Department of 
Energy by Lawrence Livermore National Laboratory under contract No. 
DE-AC52-07NA27344. 

The National Radio Astronomy Observatory is a facility of the National 
Science Foundation operated under cooperative agreement by Associated 
Universities, Inc.

\clearpage

\appendix

\section{Appendix: Notes on Individual Objects}

\subsection{F2M0729+3336}

This object lies outside the SDSS coverage, so photometry from the ESI 
spectrum was used when constructing the SED. Both the silicate feature and 
the optical SED are well fit with a reddening of $E(B-V)=0.83\pm 0.22$. The 
object is detected in the X-ray by {\em Chandra} with six hard X-ray 
photons \citep{redqso-x}. It has an accretion rate close to (formally 
slightly in excess of) the Eddington limit, but no evidence of cool dust from 
substantial star formation activity, though the host galaxy appears 
disturbed. The inferred black hole mass lies significantly below the 
black hole mass -- bulge luminosity relation, consistent with our observation 
that the black hole is still growing rapidly.

\subsection{F2M0825+4716}

Another high accretion rate object, but this one also has a strong starburst 
component in the far-infrared. The host is very disturbed, but, if the fit to 
the central component of the galaxy only is considered, lies close to the 
black hole mass -- bulge luminosity relation. The strong black hole growth 
would thus take it significantly above the relation within a Salpeter time, 
if it were not for the high star formation rate, and the likelihood that some 
of the extended components of the host galaxy would merge with the central 
one in that timescale. As such, F2M0825+4816 is a prime example of black 
hole -- host galaxy co-evolution.

\subsection{F2M0830+3759}

The lowest redshift (z=0.414) object in our sample, with a strong Chandra 
detection \citep{redqso-x}. A deeper X-ray spectrum with XMM reveals a cold 
or only moderately ionized absorber with a column density consistent with it
 being associated with the dust causing the quasar reddening, as well as an 
Fe K$\alpha$ emission line from a reflection component \citep{picon10}. This 
object has a relatively massive host galaxy, and the object lies below the 
black hole mass -- bulge luminosity relation. Nevertheless, the accretion 
rate is relatively high $L/L_{\rm Edd} \approx 0.4$, so again, this quasar 
may be able to build black hole mass effectively. The star formation rate is 
constrained to be relatively low, $\sim 10M_{\odot}{\rm yr^{-1}}$.

\subsection{F2M0834+3506}

This object has little or no silicate absorption, inconsistent with the 
observed reddening in the optical. The HST/ACS image too suggests that this, 
rather than being a post-merger system, may be a normal quasar obscured by a 
foreground irregular galaxy. The spectrum shows multiple-peaked permitted 
lines, with three identifiable components, one at the same redshift as the 
forbidden lines, one at $\approx +3600{\rm kms^{-1}}$ and one at $\approx 
-5100{\rm kms^{-1}}$. The object is the only one of two in our sample to show 
multiple-peaked permitted lines, typically only seen in low accretion rate, 
radio-loud objects \citep{eracleous03}, and our analysis indeed confirms a 
low accretion rate and a relatively low star formation rate. It lies a little 
above the local black hole mass -- bulge luminosity relation, consistent with 
the idea that this object is a fairly normal quasar obscured by dust from a 
foreground object.

\subsection{F2M0841+3604}

Consistent the very irregular morphology, this object is well-fit by the 
simple cold dust screen model. It has both a moderately-high accretion rate 
and moderately-high star formation rate, lying below the black hole mass -- 
bulge luminosity relation. The radio information reveal a steep spectrum 
point source, while the X-ray image shows a very hard hardness ratio 
\citep{redqso-x}.

\subsection{F2M0915+2418}

This object is the most luminous quasar in our sample, and is bright enough 
in the infrared to be detected in the IRAS Faint Source Catalog. The IRS 
spectrum is an archival one from \cite{sargsyan08}. This object has both a 
high star formation rate and high accretion rate, and is the only one of our 
high accretion rate objects to fall close to the black hole mass -- bulge 
mass relation in both plots of Figure \ref{mbhlstar}. Due to its high star 
formation rate and high accretion rate, it will evolve with a trajectory 
close to the black hole mass -- bulge mass relation for at least a 
Salpeter time ($\sim 10^8$yr).

\subsection{F2M1012+2825}

A fairly unexceptional object in terms of its accretion rate, star formation 
rate and position on the black hole mass -- bulge luminosity relation, it 
does, however have a double nucleus with a separation of only 1.2kpc 
\citep{redqso-hst}.

\subsection{F2M1113+1244}

Another good candidate for a young quasar, with a high accretion rate (super 
Eddington; $log(L/L_{\rm Edd}) = 0.4 \pm 0.1$), high star formation rate (FIR 
contribution in ULIRG regime; $log(L_{\rm FIR}/L_\odot)$ = 12.39), and 
plotting well below the black hole mass -- bulge mass relation.

\subsection{F2M1118-0033}

This object shows no clear broad lines in either the optical or near-infrared 
spectrum. H$\alpha$ is detected, but is relatively narrow ($\approx 
1000{\rm kms^{-1}}$). There is a hint of broad Pa$\beta$ in the near-infrared 
$K$-band, but it is not strong enough to allow us to estimate a linewidth. 
The SED is consistent with a reddened quasar. No black hole mass estimate is 
possible in this case, though we have obtained a limit based on the quasar 
luminosity and assuming Eddington-limited accretion.

\subsection{F2M1151+5359}

The reddening of this quasar is relatively modest, and the host almost 
undisturbed. Nevertheless, it has a high accretion rate and relatively high 
star formation rate, and lies well below the black hole mass -- bulge mass 
relation.

\subsection{F2M1507+3129}

\begin{figure}
\plotone{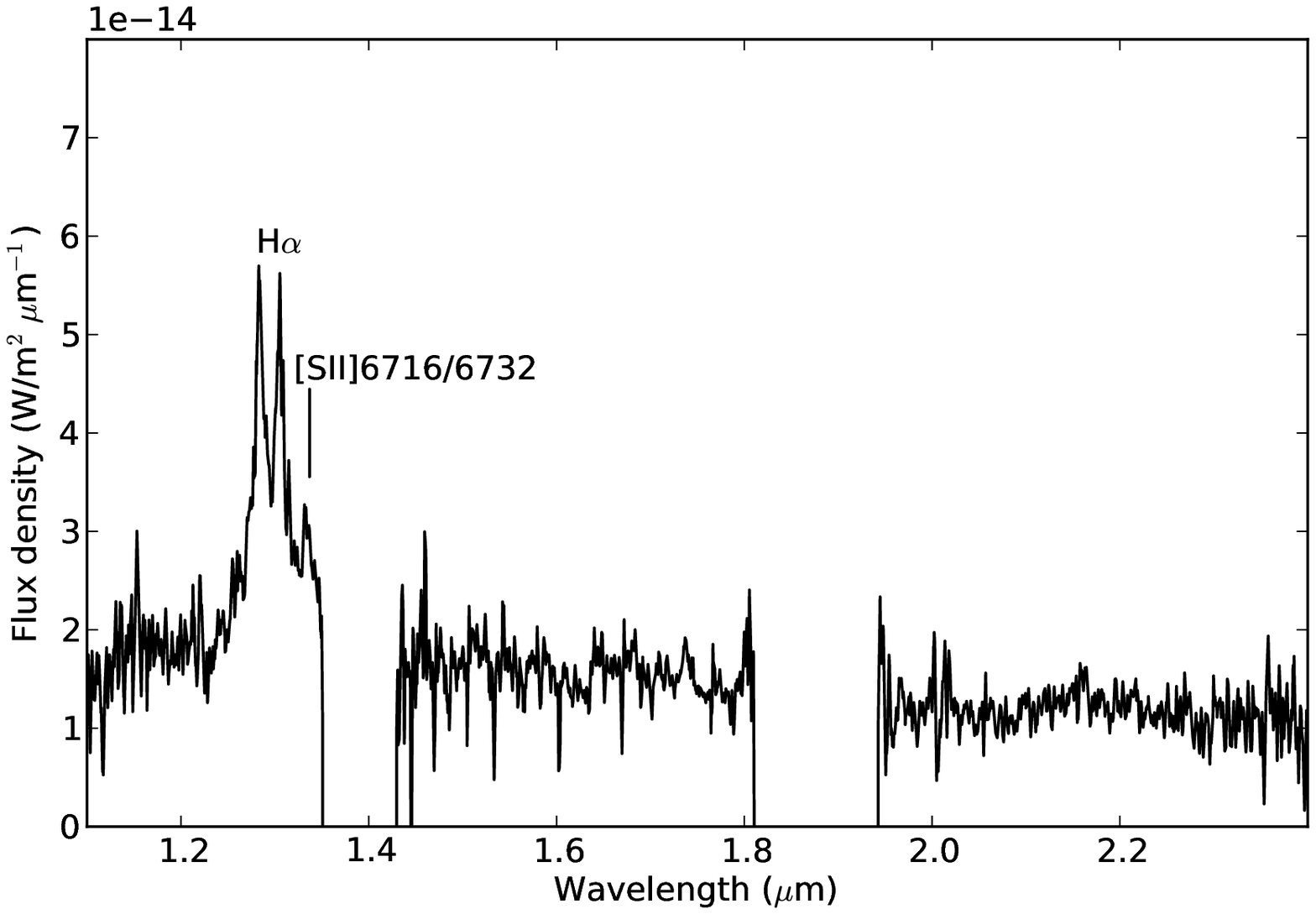}
\label{f2m1507spec}
\caption{Near-infrared spectrum of F2M1507+3129, showing a double-peaked 
H$\alpha$ line.}
\end{figure}

This object has a near-infrared spectrum in \cite{eilat04}, however, we 
obtained an improved near-infrared spectrum with the Triplespec instrument 
\citep{herter08} on the Hale 200-inch telescope on July 11th 2011. The data 
were calibrated using the A0V star HD127304. The new spectrum clearly shows a 
double-peaked broad H$\alpha$ emission line in $J$-band (Figure 
\ref{f2m1507spec}). The object is the second highest redshift double-peaked 
quasar currently known, and is more luminous than CXOECDFSJ0331-2755 at 
$z=1.369$, the most distant double-peaked emitter \citep{luo09}. Like many 
objects in this class \citep{lh06}, and indeed similar to the other 
double-peaked emitter in our sample, F2M0834+3506, it has a relatively 
low accretion rate (0.05 Eddington). 

\subsection{F2M1532+2415}

The SED of this object is difficult to fit, consistent with the complicated 
structure seen in the host galaxy image. Despite a red continuum, the IRS 
spectrum lacks silicate absorption. The far-infrared luminosity is very 
low, the accretion rate is also low and the object lies close to the black 
hole mass -- bulge mass relation. As of now the underlying physics 
responsible for the reddening and host disturbance is difficult to explain.

\subsection{F2M1656+3821}

This object has a relatively low star formation rate (barely above the LIRG 
regime; ${\rm log}(L_{\rm FIR}/L_\odot)$ = 11.19), but a high accretion rate 
(roughly Eddington rate), and should be able to accrete matter onto the black 
hole to move up to the black hole mass -- bulge mass relation from its 
current position $\approx 0.5$ dex below. It is also the object that is 
clearly in quadrant 2A of the \cite{spoon07} diagram, since it has such a 
deep Silicate absorption feature.

\end{document}